\documentclass[twocolumn]{aastex631}

\usepackage{xcolor}
\usepackage{CJK}

\newcommand{\zn}{PKS~0745-191}
\newcommand{\nii}{[{\sc N\,ii}]}
\newcommand{\sii}{[{\sc S\,ii}]}

\newcommand{\chandra}{\textit{Chandra X-ray Observatory}}

\begin{document}

\title{HST Observations within the Sphere of Influence of the Powerful Supermassive Black Hole in \zn\/}

\shorttitle{HST Observations of \zn\/}
\shortauthors{Hlavacek-Larrondo et al.}

\begin{CJK}{UTF8}{}
\CJKfamily{mj}

\correspondingauthor{Julie Hlavacek-Larrondo}
\email{j.larrondo@umontreal.ca}

\author[0000-0001-7271-7340]{Julie Hlavacek-Larrondo}
\affiliation{D\'{e}partement de Physique, Universit\'{e} de Montr\'{e}al, Succ. Centre-Ville,
Montr\'{e}al, Qu\'{e}bec, H3C 3J7, Canada}

\author[0000-0002-3173-1098]{Hyunseop Choi (최현섭)}
\affiliation{D\'{e}partement de Physique, Universit\'{e} de Montr\'{e}al, Succ. Centre-Ville,
Montr\'{e}al, Qu\'{e}bec, H3C 3J7, Canada}

\author[0000-0002-3680-5420]{Minghao Guo (郭明浩)}
\affiliation{Department of Astrophysical Sciences, Princeton University, Princeton, NJ 08540, USA}

\author[0000-0001-7597-270X]{Annabelle Richard-Laferri\`{e}re}
\affiliation{Institute of Astronomy, University of Cambridge, Madingley Road, Cambridge CB3 0HA, UK}

\author{Carter Rhea}
\affiliation{D\'{e}partement de Physique, Universit\'{e} de Montr\'{e}al, Succ. Centre-Ville,
Montr\'{e}al, Qu\'{e}bec, H3C 3J7, Canada}

\author{Marine Prunier}
\affiliation{D\'{e}partement de Physique, Universit\'{e} de Montr\'{e}al, Succ. Centre-Ville,
Montr\'{e}al, Qu\'{e}bec, H3C 3J7, Canada}

\author{Helen Russell}
\affiliation{School of Physics \& Astronomy, University of Nottingham, University Park, Nottingham NG7 2RD, UK}

\author{Andy Fabian}
\affiliation{Institute of Astronomy, University of Cambridge, Madingley Road, Cambridge CB3 0HA, UK}

\author{Jonelle L. Walsh}
\affiliation{Department of Physics and Astronomy, Texas A\&M University, College Station, TX, 77843-4242 USA}
\affiliation{George P. and Cynthia Woods Mitchell Institute for Fundamental Physics and Astronomy, Texas A\&M University, TX, 77843-4242 USA}

\author{Marie-Jo\"{e}lle Gingras}
\affiliation{Department of Physics and Astronomy, University of Waterloo, 200 University Avenue West, Waterloo, ON N2L 3G1, Canada}
\affiliation{Waterloo Centre for Astrophysics, Waterloo, ON N2L 3G1, Canada}

\author{Brian McNamara}
\affiliation{Department of Physics and Astronomy, University of Waterloo, 200 University Avenue West, Waterloo, ON N2L 3G1, Canada}
\affiliation{Waterloo Centre for Astrophysics, Waterloo, ON N2L 3G1, Canada}

\author{Steve Allen}
\affiliation{Department of Physics, Stanford University, 382 Via Pueblo Mall, Stanford, CA 94305, USA}
\affiliation{Kavli Institute for Particle Astrophysics and Cosmology, Stanford University, 452 Lomita Mall, Stanford, CA 94305, USA}
\affiliation{SLAC National Accelerator Laboratory, 2575 Sand Hill Road, Menlo Park, CA 94025, USA}

\author{Andr\'{e}-Nicolas Chen\'{e}}
\affiliation{NSF NOIRLab, 670 N. A'ohoku Place, Hilo, Hawai'i, 96720, USA}

\author{Alastair Edge}
\affiliation{Centre for Extragalactic Astronomy, Department of Physics, Durham University, Durham DH1 3LE, UK}

\author{Marie-Lou Gendron-Marsolais}
\affiliation{D\'{e}partement de physique, de g\'{e}nie physique et d'optique, Universit\'{e} Laval, Qu\'{e}bec (QC), G1V 0A6, Canada}


\author{Michael McDonald}
\affiliation{Kavli Institute for Astrophysics and Space Research, MIT,
77 Massachusetts Avenue, Cambridge, MA 02139, USA}

\author{Priyamvada Natarajan}
\affiliation{Department of Astronomy, Yale University, 219 Prospect Street, New Haven, CT 06511, USA}
\affiliation{Department of Physics, Yale University, 217 Prospect Street, New Haven, CT 06511, USA}
\affiliation{Black Hole Initiative at Harvard University, 20 Garden Street, Cambridge, MA 02138, USA}

\author{Jeremy Sanders}
\affiliation{Max-Planck-Institut fur extraterrestrische Physik, Giessenbachstrasse 1, 85748 Garching, Germany}

\author{James F. Steiner}
\affiliation{Center for Astrophysics, Harvard \& Smithsonian, 60 Garden St, Cambridge, MA 02138, USA}

\author{Benjamin Vigneron}
\affiliation{D\'{e}partement de Physique, Universit\'{e} de Montr\'{e}al, Succ. Centre-Ville,
Montr\'{e}al, Qu\'{e}bec, H3C 3J7, Canada}

\author{Anja von der Linden}
\affiliation{Department of Physics and Astronomy, Stony Brook University, Stony Brook, NY 11794, USA}









\begin{abstract}

\noindent We present Space Telescope Imaging Spectrograph observations from the Hubble Space Telescope of the supermassive black hole (SMBH) at the center of \zn, a brightest cluster galaxy (BCG) undergoing powerful radio-mode AGN feedback ($P_{\rm cav}\sim5\times10^{45}$ erg s$^{-1}$). These high-resolution data offer the first spatially resolved map of gas dynamics within a SMBH’s sphere of influence under such powerful feedback. Our results reveal the presence of highly chaotic, non-rotational ionized gas flows on sub-kpc scales, in contrast to the more coherent flows observed on larger scales. While radio-mode feedback effectively thermalizes hot gas in galaxy clusters on kiloparsec scales, within the core, the hot gas flow may decouple, leading to a reduction in angular momentum and supplying ionized gas through cooling, which could enhance accretion onto the SMBH. This process could, in turn, lead to a self-regulating feedback loop. Compared to other BCGs with weaker radio-mode feedback, where rotation is more stable, intense feedback may lead to more chaotic flows, indicating a stronger coupling between jet activity and gas dynamics. Additionally, we observe a sharp increase in velocity dispersion near the nucleus, consistent with a very massive $M_{\rm BH}\sim1.5\times10^{10} M_\odot$ SMBH. The density profile of the ionized gas is also notably flat, paralleling the profiles observed in X-ray gas around galaxies where the Bondi radius is resolved. These results provide valuable insights into the complex mechanisms driving galaxy evolution, highlighting the intricate relationship between SMBH fueling and AGN feedback within the host galaxy.
\end{abstract}


\section{Introduction} \label{sec:intro}

Supermassive black holes (SMBHs) play a crucial role in the formation and evolution of galaxies. These colossal entities, residing at the centers of most large galaxies, influence their host galaxies through processes known as active galactic nucleus (AGN) feedback. AGN feedback regulates star formation by heating and expelling gas, thereby preventing it from cooling and collapsing to form new stars. Additionally, this feedback drives galactic winds and outflows that redistribute gas and metals, thus shaping the galaxy's structure and evolution over cosmic time. Understanding AGN feedback is crucial to explain observed galaxy properties, such as the correlation between SMBH mass and stellar velocity dispersion, as well as the overall energy balance within galaxies \citep[see reviews by][]{Kormendy2013,Fabian2012,McNamara2012}.

AGN feedback can generally be divided into two distinct categories, each significantly impacting the host galaxy in different ways. The first occurs when the SMBH is accreting at high rates, known as the radiatively-efficient mode (also referred to as quasar mode). Here, SMBHs typically accrete at rates above 1 percent of the Eddington rate. The accretion disk in such cases is expected to be geometrically thin (though it may become thick at very high accretion rates), and most of the feedback emerges in the form of radiation and powerful winds \citep[e.g.,][]{King2015,Hopkins2010}. The second mode, which will be the focus of this paper, occurs when the SMBH is accreting at low rates, typically well below 1 percent of the Eddington rate. This mode, known as the radiatively-inefficient mode (also referred to as radio-mode, jet-mode, maintenance-mode, mechanical-mode, or kinetic-mode), is potentially the dominant one in terms of total energy injected into the galaxy throughout its lifetime \citep{Heckman2023}. Here, most of the feedback emerges in the form of powerful radio jets that heat the surrounding gas, redistribute metals throughout the host galaxy and beyond, and create powerful molecular outflows that completely reshape the properties of the host galaxy \citep[see reviews by][]{Fabian2012,McNamara2012}.

Radio-mode feedback is best studied in the most massive galaxies, located at the centers of galaxy clusters. This is because the surrounding medium, known as the intracluster medium (ICM), shines brightly in X-rays, allowing us to directly observe the impact of radio jets on this gas. Here, the SMBH located in the central dominant galaxy drives powerful relativistic jets that carve out gigantic X-ray cavities which prevent most of the ICM from cooling, thereby reducing the expected star formation rates by orders of magnitude \citep{McNamara2012,Fabian2012}. This feedback appears to be remarkably self-regulated, which is puzzling given that the size of a black hole is typically $10^6$-$10^9$ times smaller than its host galaxy.

\begin{figure*}
\centering
\includegraphics[width = 1\textwidth]{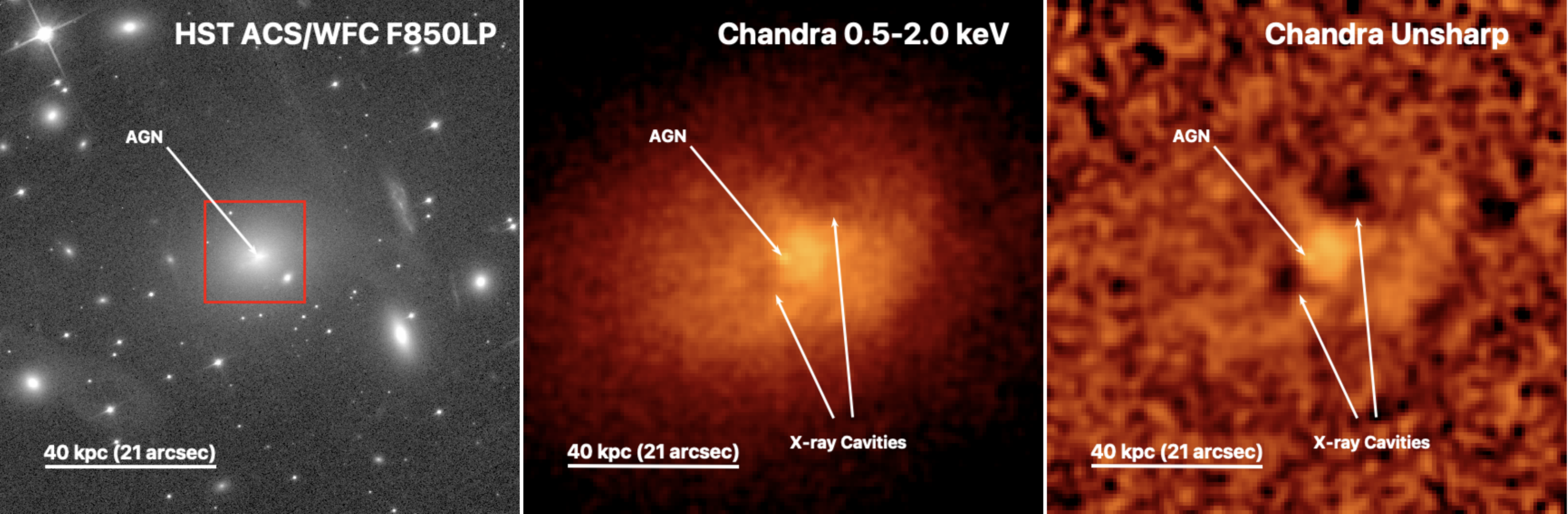}
\caption{Image of the central 150 kpc $\times$ 150 kpc region in the massive cluster of galaxies \zn\/ (same field of view in all panels). North is up and East is left. Left: HST F850LP image, with location of the central AGN in the BCG highlighted. The red box illustrates the region shown in Fig. \ref{fig:mediumscale}. Middle: $Chandra$ 0.5-2.0 keV X-ray image of the cluster. The image has been smoothed with a gaussian filter of $\sigma$=2 pixels. The location of the central AGN and two X-ray cavities are shown. Right: Unsharp masked $Chandra$ 0.5-2.0 keV image produced by smoothing the original image by a gaussian filter of $\sigma$=1 pixel and subtracting this by a smoothed image using a gaussian filter of $\sigma$=8 pixels. This image clearly reveals the X-ray cavities.}
\label{fig:largescale}
\end{figure*}

\noindent To address this puzzle, a new model has emerged to explain the self-regulated connection between a SMBH and its surrounding medium. This model, often called the \textit{precipitation model} \citep[e.g., ][]{Voit2015-N}, states that the circumgalactic medium, like the ICM, when cooled, precipitates and condenses into cooler clouds. These clouds eventually rain down onto the central black hole, causing it to re-ignite and heat the environment. Similar models include \textit{cold feedback} \citep[e.g., ][]{Pizzolato2005}, \textit{chaotic cold accretion} \citep[e.g., ][]{Gaspari2013}, and \textit{stimulated feedback} \citep[e.g., ][]{McNamara2016}.

\noindent A clear prediction of the precipitation model is that some of the infalling clouds must have low specific angular momentum to reach the black hole \citep{Pounds2018}. The model, therefore, predicts that condensation should be \textit{chaotic}, especially near the black hole \citep[e.g.,][]{Gaspari2018}. To date, the most direct observational evidence supporting this type of chaotic accretion has been the detection of cold molecular gas through absorption line features using the galaxy's own radio core as a backlight \citep[e.g.,][]{David2014,Tremblay2016,Rose2019a,Rose2019b}. This cold gas appears to be within the inner few hundreds of parsecs of the galaxy's SMBH and is consistent with inflow \citep[e.g.,][]{Rose2019b}. However, the main drawback with this kind of spectroscopic data is that it only provides single pencil-beam sight-line of the accretion process (i.e., only radial velocity along one sight-line). Consequently, it remains unclear whether the flow is chaotic beyond these limited sight-lines.

In this study, we focus on the nearest cluster of galaxies with a mass deposition rate greater than 1000 M$_{\odot}$ yr$^{-1}$ \citep[][ and references therein]{Fabian1985}, exhibiting some of the most powerful radio-mode feedback observed in any cluster, with $P_{\rm cav}\sim5\times10^{45}$ erg s$^{-1}$ \citep[][]{Rafferty2006}. Our goal is to provide the first resolved map of the gas flow within the sphere of influence of the central SMBH located in the Brightest Cluster Galaxy (BCG) using the Space Telescope Imaging Spectrograph (STIS) on the \textit{Hubble Space Telescope} (\textit{HST}).

\zn\/, located at a redshift of $z=0.102428$ \citep[see Fig. \ref{fig:largescale} here, as well as][]{Marie-Joelle}, hosts a powerful radio source in its central dominant galaxy, ranking among the top 5 percent hosted in such cluster galaxies \cite{Hogan2015}. \citet{Baum1991} conducted multi-frequency VLA observations of \zn\/, revealing double-sided radio jets oriented in the East-West direction at 15 GHz, spanning approximately 1\arcsec ($\sim2$ kpc) in length. Beyond this compact jet structure, a more amorphous, diffuse morphology extending to scales of about 7-14\arcsec ($\sim13-25$ kpc) was also observed. This complex radio morphology distinguishes \zn\/ from the conventional Fanaroff-Riley (FR) I and II classifications commonly associated with active radio-mode feedback in galaxy clusters.

\zn\/ is also one of the most massive clusters of galaxies known, with $M_{500}= (7.3 \pm 0.8) \times 10^{14}~$M$_{\odot}$ \citep[][]{Arnaud2005}. Additionally, it hosts a massive reservoir of cold molecular gas, totalling several $10^9$ M$_\odot$ \citep{Salome2003,Russell2016}. On large kpc-scales, the flow of the cold molecular gas appears inconsistent with a merger origin or gravitational free-fall of cooling gas \citep{Russell2016}. Instead, comparisons with the X-ray cavities formed by the radio jets indicate that the molecular gas is being formed in the updraft behind the X-ray cavities rising through the ICM of the cluster. \citet{Wilman2009} also used \textit{VLT-SINFONI} to study the kinematics of ionized and molecular gas in \zn\/ on kpc-scales, identifying extended line emission features ($\sim 5$ kpc) and gradual velocity shifts. More recently, \citet{Marie-Joelle} observed \zn\/ with the Keck Cosmic Web Imager, detecting significantly larger nebular emission features ($\sim 10-20$ kpc) traced by [\ion{O}{2}] emission. These observations revealed disrupted gas and high-velocity dispersions near the nucleus, likely driven by AGN activity, suggesting that radio-mode AGN feedback plays a role in disturbing the gas on larger scales. Despite the powerful feedback, \zn\/ also appears to be fuelling a star formation rate of $\sim20$ M$_\odot$ yr$^{-1}$ \citep[e.g.,][]{Mittal2015}, as well as a filamentary nebula of ionized gas \citep[e.g.,][]{Fabian1985,Johnstone1987,DOnahue2000,Hicks2005,Tremblay2015}. Its ICM has also been extensively studied out to very large radii beyond $r_{200}$, suggesting that the ICM is out of hydrostatic equilibrium \citep[][]{Walker2012}.

\begin{figure*}
\centering
\includegraphics[width = 1\textwidth]{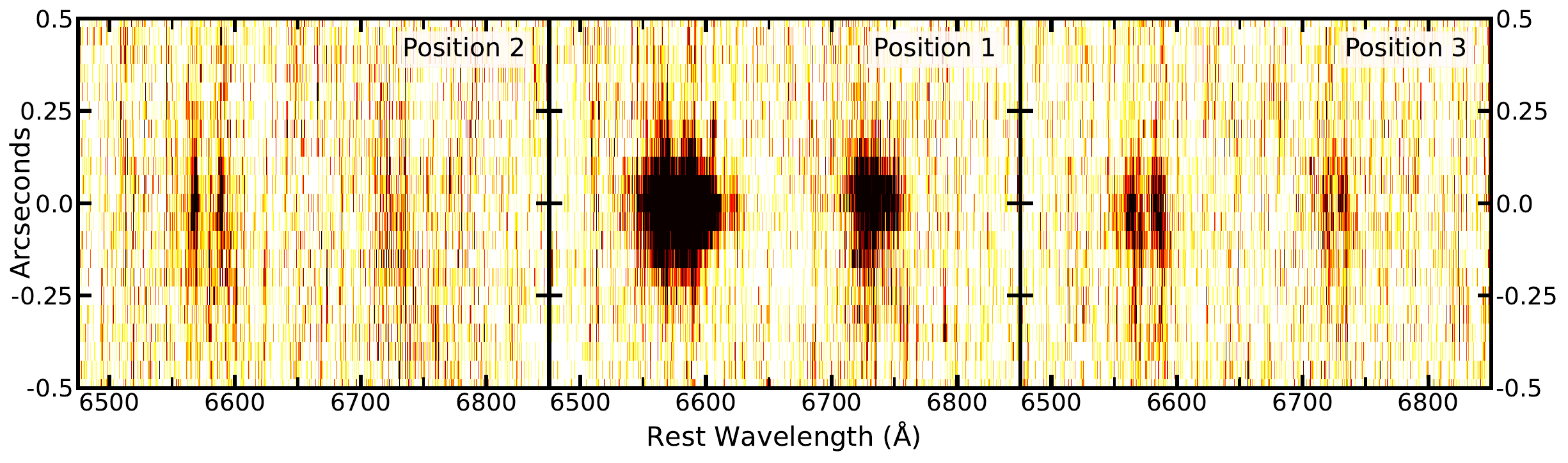}
\caption{The 2D continuum subtracted images extracted from STIS G750M long-slit observations at three slit positions centered around \zn. The locations of the slit are reported in the panels, each offset by 0.1\arcsec from the central Position 1.
The line complexes seen are from \nii{}$\lambda$6548, \nii{}$\lambda$6584, H$_\alpha\lambda$6563, \sii{}$\lambda$6716 and \sii{}$\lambda$6731 emission.}
\label{fig:2d_spec_img}
\end{figure*}

The aim of this paper is to present the first kinematical observations within the sphere of influence of a SMBH undergoing powerful radio-mode feedback, with the objective of gaining insight into the gas dynamics surrounding the SMBH during this accretion phase. In Section \ref{sec:obs}, we present the new observations acquired using HST with STIS, the Advanced Camera for Surveys (ACS), and the Wide Field Camera 3 (WFC3). 
In Section \ref{sec:ana}, we present the analysis of the STIS data. In Section \ref{sec:res}, we show the results, and in Section \ref{sec:disc}, we discuss their implications. Throughout this paper, we assume $H_0=70$ km s$^{-1}$ Mpc$^{-1}$, $\Omega_{\rm m}=0.3$, and $\Omega_{\rm \Lambda}=0.7$. We also assume a redshift of $z=0.102428$ as the reference redshift, which was determined from stellar population models \citep{Marie-Joelle}. This corresponds to 1.884 kpc per \arcsec. All errors are 1$\sigma$ unless otherwise specified.

\section{Observations} \label{sec:obs}

\subsection{Hubble Space Telescope} \label{sec:obsHST}

 \subsubsection{STIS} 
 
The \textit{HST} STIS observations of \zn\/ were made on 2016 November 4--5 under the program GO-14669 (PI Hlavacek-Larrondo). We designed the observations to obtain the spatial maps of the kinematic and physical properties of the nebular emission lines around the central SMBH.
We used the G750M grating, which covered the required wavelength range needed to trace the \nii{}$\lambda$6548, \nii{}$\lambda$6584, H$_\alpha\lambda$6563, \sii{}$\lambda$6716 and \sii{}$\lambda$6731 kinematics, while having a dispersion of 0.56 Angstrom per pixel with a Full Width Half Maximum (FWHM) of $\sim45$ km s$^{-1}$ at H$_\alpha$).
The CCD readout was performed in an unbinned mode with a wavelength scale of 0.55 $\rm \AA\ pixel^{-1}$ and a spatial scale of $0.05\arcsec$ pixel$^{-1}$.
We required a minimum signal-to-noise ratio (SNR) of 2 per resolution element \citep[same as][]{DallaBonta2009}, resulting in a line SNR of 10.

In order to accurately probe the flow within the sphere of influence, we used 3 different slit positions. We used a slit width of 0.1\arcsec\ by 52\arcsec, which allowed us to resolve the sphere of influence while allowing for optimum throughput. To optimize the cost time of the observations and to spectroscopically map the region, we obtained 1 set of observations in 3 dithered positions, such that one was on the nucleus (RA 07:47:31.32, Dec. -19:17:39.97, J2000), and two others were offset by $\pm0.1$\arcsec. The dither pattern was set to STIS- PERP-TO-SLIT combined with the secondary STIS-ALONG-SLIT pattern to further improve the image processing (e.g., to dither hot pixels). The total area covered was then 0.3\arcsec\ by 52\arcsec. We refer to the central slit location, Position 1, and Positions 2 and 3 correspond to the slit offset by 0.1\arcsec\ towards east and west relative to the target, respectively.

\begin{figure*}
\centering
\includegraphics[width = 1\textwidth]{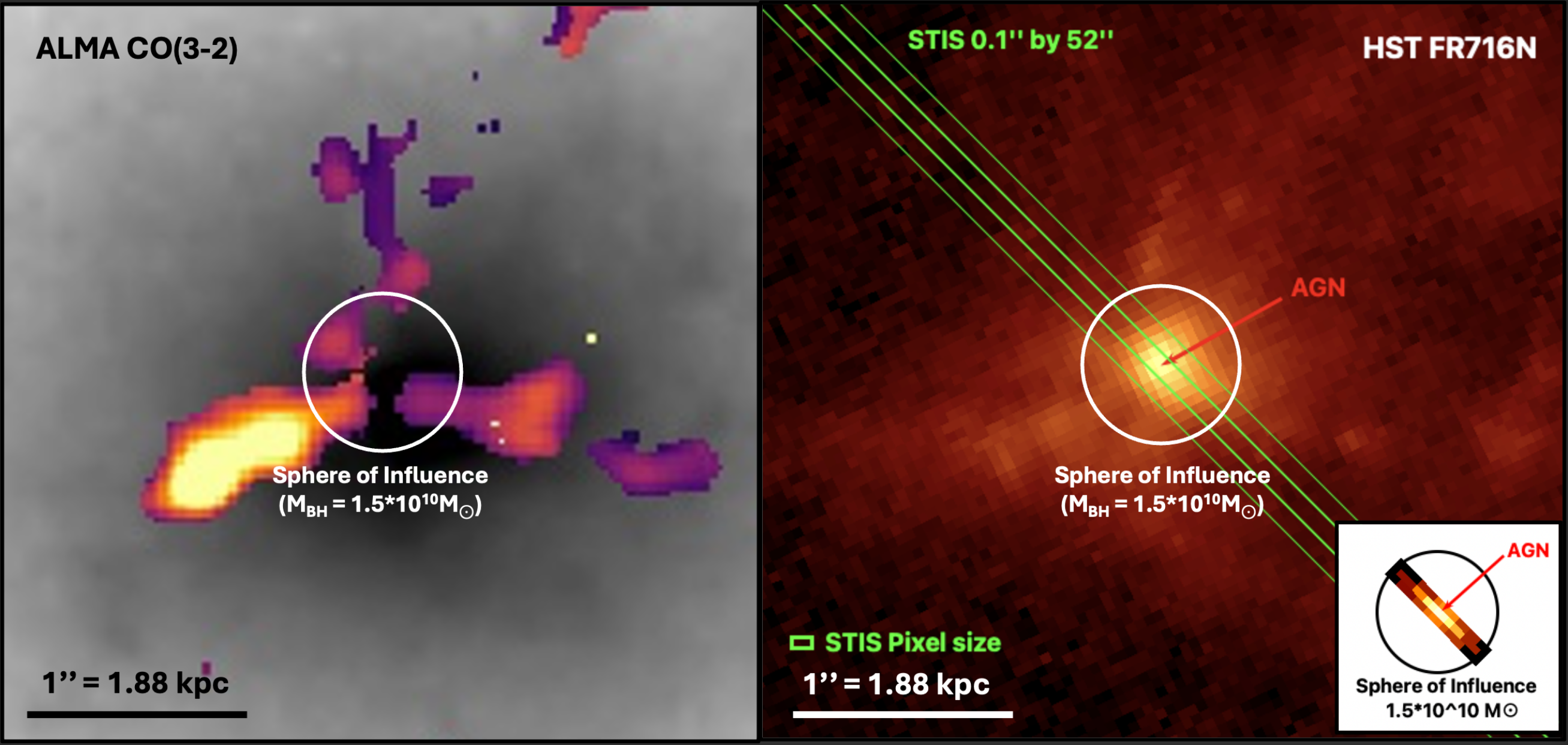}
\caption{Zoom-in onto the BCG in \zn\/ in a 4\arcsec by 4 \arcsec box. Left: ALMA CO(3-2) flux map showing a lack of emission in the central region. Right: ACS FR716N narrow-band image that covers H$\alpha$/[\ion{N}{2}] emission lines. We highlight the location of the three STIS slits, each with a size of 0.1\arcsec by 52\arcsec, as well as the size of each pixel along the slit. In addition, we show the sphere of influence for a $1.5\times10^{10}$ M$_\odot$ SMBH. Right-inset: Zoom-in on the H$\alpha$ flux map measured from the STIS observations, with the sphere of influence overplotted.   }
\label{fig:zoom}
\end{figure*}

The raw STIS observation data were retrieved from the Mikulski Archive (MAST) and reduced in a standard manner. We trimmed the overscan region and subtracted the bias and dark from each exposure and performed flat-field corrections. Once this processing was done, we proceeded with cosmic ray rejections and hot pixel removals using \texttt{LA-COSMIC} \citep{vanDokkum2001}.
We visually inspected the corrected images for both science and calibration exposures (e.g., wavecal images) for the three slit positions to verify their quality.
We used the standard Space Telescope Science Institute (STScI) pipeline in \texttt{IRAF} to complete the reduction process. The resulting reduced and geometrically rectified exposures from each slit position were combined using the \texttt{IRAF} tasks \texttt{IMSHIFT} and \texttt{IMCOMBINE} to produce the final 2D spectral images. The total exposure time at each slit position with the dither patterns was 9600 seconds.
The STIS slits were set to be oriented at a position angle (P.A.) of $-135^{\circ}$ ($45^{\circ}$ from North towards East) in order to best follow the main axis of the large-scale kinematics of \citet{Wilman2009}. This also allowed us to map the kinematics perpendicular to the X-ray cavities (see Fig. \ref{fig:zoom}). Throughout the paper, we use a positive sign to refer to the offset locations along the slit length toward North-East.

Figure \ref{fig:2d_spec_img} shows the 2D continuum subtracted images extracted from the STIS  data along the 3 slit positions centered around \zn.

\begin{figure*}
\centering
\includegraphics[width = 1\textwidth]{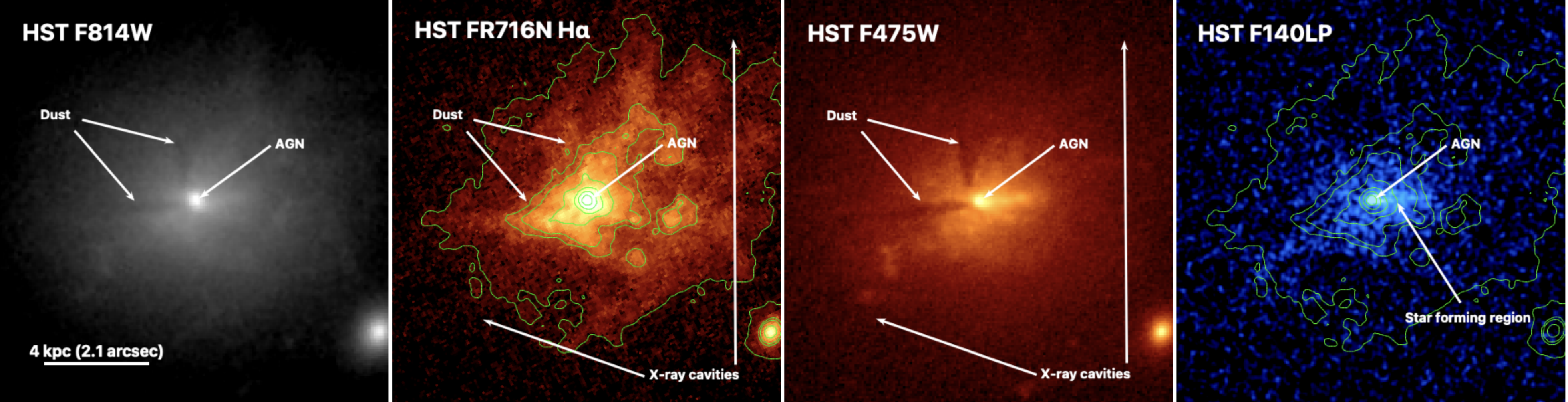}
\caption{HST images of the BCG in \zn\/. Left: WFPC2 F814W filter tracing the older stellar population, including prominent dust lanes. Middle-left: ACS FR716N narrow-band image that spans the wavelength range covering the H$\alpha$/[\ion{N}{2}] emission lines. Contours start at 3$\sigma$ and increase by factors of two. We highlight the location of the X-ray cavities as seen on the right panel of Figure \ref{fig:largescale}. Middle-right: ACS F475W image tracing the B-band emission. Right: SBC F140LP image tracing the star formation ($\sim20$ M$_\odot$ yr$^{-1}$) with FR716N narrow-band image contours overlaid. We highlight the location of the AGN and the star forming region. The H$\alpha$/[\ion{N}{2}] emission trail behind the X-ray cavities as in Fig. \ref{fig:largescale}.}
\label{fig:mediumscale}
\end{figure*}

\subsubsection{HST imaging}

The new \textit{HST} ACS and WFC3 observations were conducted as part of the \textit{HST} STIS program, and the exposures were taken on 2017 January 30.
We retrieved the fully reduced, science-ready images taken with the ACS and the WFC3 using the filters FR716N, F850LP, and F475W from MAST. The FR716N filter extracts the total emission on large-scales from \nii{}$\lambda$6548, \nii{}$\lambda$6584 and H$_\alpha\lambda$6563. The F850LP image traces the stellar continuum emission and the F475W traces the B-band emission. 

The total exposure times for FR716N, F850LP, and F475W images were 1050, 1314, and 1732 seconds. Upon evaluating the images, we noticed that the standard MAST processing pipeline was not adequate for the narrow-band FR716N filter image. It prominently showed the post-Servicing Mission 4 row-correlated ``stripping noise'', also called bias striping.
Therefore, we downloaded the raw data and performed calibrations with extra steps to clean the detector artifacts. We included additional steps using tasks in the ACS destripping tool\footnote{https://acstools.readthedocs.io/en/latest/acs$\_$destripe.html} provided by the STScI \citep{Grogin2011} to remove the image artifacts from each exposure taken with FR716N.
The destriped images were re-aligned and combined/drizzled using the tasks included in the \texttt{DrizzlePac} software package \citep{Fruchter2010}. Here, we do not remove the continuum from the image as we also show it for illustrative purposes. 



Additionally, we retrieved archival HST/ACS image taken during the GO-12220 (P.I. Mittal, R.) using the Solar Blind Channel (SBC) and WFPC2 images taken during the GO-7337 (P.I. Fabian, A.).
These images include observations conducted with the long-pass UV filter F140LP for the SBC and the broadband filer F814W for the WFPC2. The F140LP reveals where the star formation is occurring and the F814W traces the old stellar population and includes prominent dust features in the BCG. Here, we used the images produced by MAST as part of the Single Visit Mosaics (SVMs) data products from the Hubble Advanced Products (HAP) reprocessing.

We show the resulting HST images of the cluster in Fig. \ref{fig:largescale} and of the BCG in Fig. \ref{fig:mediumscale}. In Fig. \ref{fig:zoom}, we also show a Zoom-in onto the central regions of the BCG, along with the location of the STIS slits, and the sphere of influence of a typical $10^{10}$ M$_\odot$ SMBH.



\subsection{Chandra X-ray Observatory} \label{sec:obsChandra}

\zn\/ has been observed by the \chandra{} a dozen times since its inception \citep[see][]{Sanders2014}; here, we present the two ACIS-S observations that exhibit a steady background lightcurve (ObsID 2427 and ObsID 12881). These observations have a combined exposure rate of 136 ks. The data were processed using a modified reduction algorithm presented in \cite{Rhea2020zndo}\footnote{The code is available at \url{https://github.com/XtraAstronomy/AstronomyTools/tree/master/DataCleaning}}.
The cleaning algorithm starts by having the user select a background CCD that does not contain source emission and with the same type of chip (front versus back illuminated). Using this CCD, we create a background lightcurve, which is used to identify time intervals during which the background emission flares. These flare events are defined by a flux outside of a 3-$\sigma$ baseline calculated for the background CCD. Time intervals categorized as flaring are removed from the analysis. The data is then destreaked, cleaned of bad pixels, and processed using the \texttt{CIAO} tool \texttt{acis\_process\_events} with \texttt{VFAINT=True}. Background files are created using the \texttt{CIAO} tool \texttt{blanksky}.
This process is repeated for both ObsIDs independently. The final X-ray image in the 0.5 to 2.0 keV band is shown in Fig. \ref{fig:largescale}. We also show the unsharp-masked image produced by smoothing the original image by a gaussian filter of $\sigma$=1 pixel and subtracting this by a smoothed image using a gaussian filter of $\sigma$=8 pixels. This image clearly reveals the large X-ray cavities in the cluster.

In order to construct a radial profile of the density of the ICM, which will be compared to the density extracted from STIS in \S~\ref{sec:res}, we constructed regions of concentric annuli centered on the AGN, which also manifests itself as the brightest X-ray point in the cluster. We created ten regions of constant signal-to-noise ratio set to 35. \footnote{The algorithm can be found at \url{https://github.com/XtraAstronomy/AstronomyTools/blob/master/Regions/CreateAnnuli.py}.} The spectra are then extracted using the \texttt{CIAO} tool \texttt{specextract}. This process creates both the source and background spectra and generates the corresponding response files. We then fitted each spectrum using an absorbed \texttt{APEC} model, where we let the temperature, abundance, and normalization parameters free to vary. From the normalization value, we extracted the electron density of the gas \citep{Calzadilla2019}.
We compare these densities to the ones extracted from HST STIS in \S~\ref{sec:disc_flow_den}. 

For the central annulus, due to the strong presence of the AGN in the center bin, we use the following {\sc mekal} model, phabs [mekal + zphabs(pow+ga)], developed in
\citet{HL2013}.
The background spectra were subtracted from the source spectra prior to the fitting procedure. We use the \texttt{cstat} to calculate the goodness of fit.
Each individual spectrum is fit using \texttt{emcee}\footnote{https://emcee.readthedocs.io/en/stable/} \citep{emcee2013}. We assume a Gaussian likelihood function.
The fitting procedure consists of two steps: 1) calculate the initial values for the random walks by applying the Neldermead algorithm as implemented in \texttt{sherpa}\footnote{http://github.com/sherpa/sherpa/;
  http://cxc.harvard.edu/sherpa/} \citep{Freeman2001} and 2) apply an \texttt{emcee} sampler for 2000 steps with 500 walkers. The first 200 steps were discarded as burnout. We assume uniform priors over the parameters of interest. The code can be found at \url{https://github.com/XtraAstronomy/AstronomyTools/blob/master/FittingPipeline/emceeFit.py}.
  
Note that we also attempted to obtain a deprojected density profile of the cluster; however, the presence of large X-ray cavities made the procedure challenging, resulting in large error bars in the profile, especially in the core, which is the region that interests us most. We therefore focus only on the projected profiles for this paper when comparing these densities to the ones extracted from HST STIS in \S~\ref{sec:disc_flow_den}.

\subsection{ALMA Observations} \label{sec:obsAlma}

The ALMA observations were obtained in Cycle 1 and presented in detail in \cite{Russell2016}. They consisted of single pointings aimed at observing both the CO(1-0) line at 104.53 GHz in band 3 and the CO(3-2) line at 313.56 GHz in band 7 (ID = 2012.1.00837.S; PI McNamara). Here, we only show the CO(3-2) observations, as these reach the highest spatial resolution, comparable to our HST STIS data. The observations were also centered on the nucleus of \zn\/, similar to our observations.

The data were initially calibrated using the ALMA pipeline reduction scripts with the Common Astronomy Software Applications (CASA) version 4.2.2 \citep{McMullin2007}, with additional self-calibration applied to the continuum data. For this paper, the data were reprocessed with uniform weighting to enhance spatial resolution, accepting a modest loss of extended emission. This approach enables a direct comparison between the STIS observations of ionized gas and the ALMA observations of molecular gas in the galaxy's core at similar spatial resolution, resulting in a synthesized beam of 0.2\arcsec\ by 0.15\arcsec. 


Spectra were extracted for synthesized beam-sized regions centered on each spatial pixel in the ALMA cube. These spectra were then fitted with a single Gaussian component using \small{MPFIT} \citep{Markwardt2009}. All lines below 3$\sigma$ using Monte Carlo simulations were excluded. The resulting flux map is presented in Fig. \ref{fig:zoom}, compared to HST FR716N image of the ionized gas, as well as the location of the STIS slits. 


\begin{figure*}
\centering
\includegraphics[width = 1\textwidth]{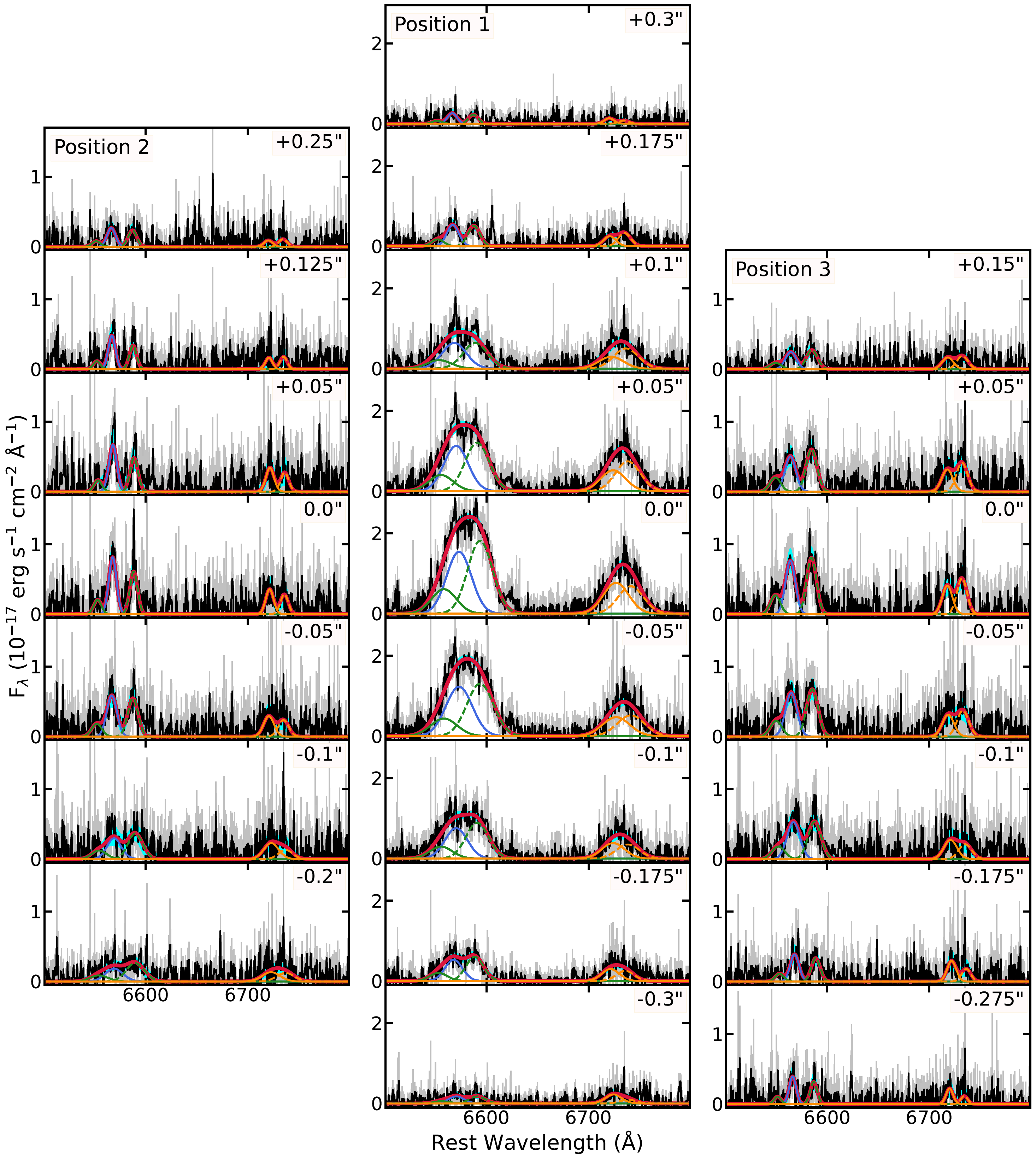}
\caption{Best-fitting models for the continuum-subtracted spectra extracted from the pixels with emission line detection at three slit locations (Positions 2, 1, and 3 from the leftmost column).
The full model that contains H$\alpha$, [\ion{N}{2}], and [\ion{S}{2}] emission lines are plotted in red, and each emission line components are plotted in blue, green, and orange, respectively (longer wavelength transitions in the doublets are shown with dashed lines).
The continuum-subtracted spectra and the uncertainties are plotted in black and grey, respectively.
We note that the pixels far from the center were binned together.
The cyan shade represents a 95\% confidence region extracted from the MCMC posterior sampling calculated using \texttt{emcee} package.
The fit results are tabulated in Table~\ref{tab:fit_table}.}
\label{fig:best_fit_all}
\end{figure*}

\begin{figure}
\centering
\includegraphics[width = 1\columnwidth]{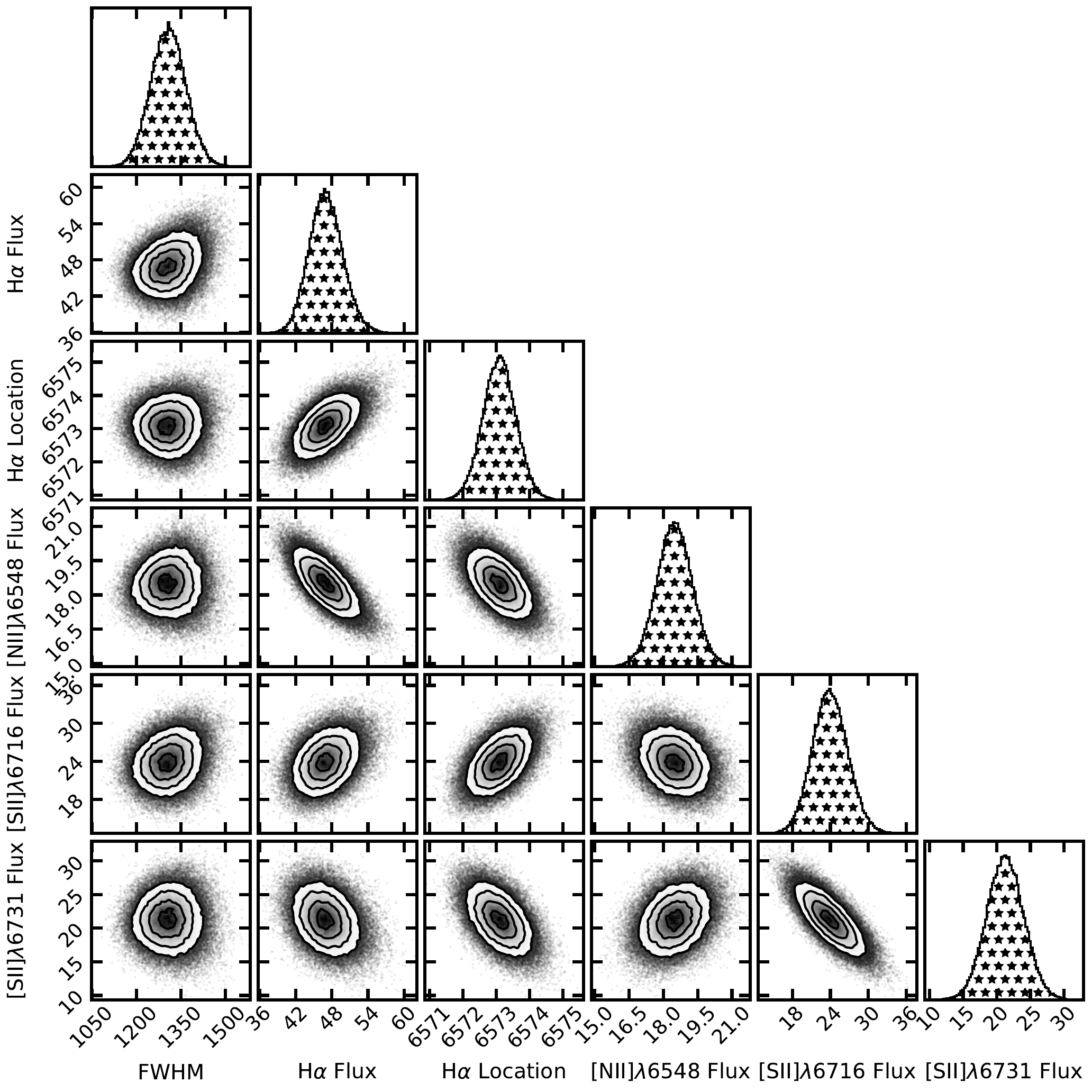}
\caption{The corner plot showing the posterior distributions of the fit parameters used in the spectral model for the centre slit (Position 1) central pixel spectrum.
We report the median values and 68\% ($1\sigma$) confidence intervals for uncertainties, unless noted otherwise.}
\label{fig:corner_central}
\end{figure}

\section{Analysis} \label{sec:ana}


\begin{deluxetable*}{CCCCCCC}[b!]
\tabletypesize{\scriptsize}
\tablecaption{Emission line fluxes and kinematics \label{tab:fit_table}}
\tablehead{\colhead{Offset} &\colhead{F(H$_\alpha$)} &\colhead{F(\nii{}$\lambda$6584)} &\colhead{F(\sii{}$\lambda$6716)} &\colhead{F(\sii{}$\lambda$6731)} &\colhead{Velocity} &\colhead{$\sigma$} \\
\colhead{(arcsec)}& \colhead{($10^{-17}\ \mathrm{erg\ s^{-1}cm^{-2}}$)} &\colhead{($10^{-17}\ \mathrm{erg\ s^{-1}cm^{-2}}$)} &\colhead{($10^{-17}\ \mathrm{erg\ s^{-1}cm^{-2}}$)} &\colhead{($10^{-17}\ \mathrm{erg\ s^{-1}cm^{-2}}$)} &\colhead{(km s$^{-1}$)} &\colhead{(km s$^{-1}$)}}
\startdata
\multicolumn{7}{c}{Position 1 (center)} \\
\hline
0.3 & 3.16^{+0.62}_{-0.6}& 2.75^{+0.5}_{-0.5}& 1.68^{+0.59}_{-0.58}& 1.0^{+0.53}_{-0.5}& 166^{+35}_{-37}& 215^{+30}_{-25} \\
0.175 & 7.96^{+0.82}_{-0.81}& 7.68^{+0.7}_{-0.69}& 3.82^{+0.91}_{-0.88}& 5.08^{+0.83}_{-0.83}& 190^{+27}_{-26}& 270^{+27}_{-22} \\
0.1 & 19.4^{+2.8}_{-2.5}& 19.3^{+2.5}_{-2.2}& 9.19^{+3.23}_{-3.36}& 15.6^{+3.7}_{-3.2}& 254^{+60}_{-62}& 562^{+68}_{-72} \\
0.05 & 33.7^{+2.7}_{-2.5}& 36.1^{+2.3}_{-2.2}& 15.9^{+2.9}_{-2.9}& 22.8^{+3.0}_{-2.9}& 338^{+32}_{-33}& 547^{+34}_{-34} \\
0.0 & 47.0^{+2.9}_{-2.7}& 55.5^{+2.3}_{-2.3}& 23.9^{+2.8}_{-2.7}& 21.3^{+2.8}_{-2.7}& 469^{+22}_{-22}& 555^{+25}_{-25} \\
-0.05 & 40.1^{+3.3}_{-3.0}& 42.7^{+2.6}_{-2.6}& 15.8^{+3.2}_{-3.1}& 17.3^{+3.1}_{-3.0}& 472^{+32}_{-31}& 596^{+34}_{-33} \\
-0.1 & 21.1^{+2.2}_{-2.1}& 25.0^{+2.0}_{-1.9}& 11.0^{+2.7}_{-2.8}& 10.0^{+2.8}_{-2.6}& 341^{+45}_{-46}& 509^{+50}_{-44} \\
-0.175 & 11.6^{+1.1}_{-1.0}& 13.2^{+1.0}_{-1.0}& 6.64^{+1.51}_{-1.47}& 5.63^{+1.36}_{-1.31}& 229^{+46}_{-42}& 394^{+47}_{-37} \\
-0.3 & 3.61^{+0.85}_{-0.8}& 3.63^{+0.81}_{-0.74}& 4.68^{+1.22}_{-1.14}& 2.0^{+1.22}_{-0.95}& 352^{+73}_{-67}& 366^{+133}_{-81} \\
\hline
\multicolumn{7}{c}{Position 2} \\
\hline
0.25 & 2.87^{+0.54}_{-0.53}& 2.53^{+0.5}_{-0.49}& 0.99^{+0.56}_{-0.53}& 1.18^{+0.52}_{-0.52}& 161^{+33}_{-32}& 193^{+25}_{-22} \\
0.125 & 4.33^{+0.74}_{-0.7}& 3.07^{+0.59}_{-0.56}& 1.51^{+0.72}_{-0.67}& 1.62^{+0.61}_{-0.6}& 185^{+22}_{-25}& 163^{+34}_{-31} \\
0.05 & 6.69^{+1.05}_{-1.0}& 4.84^{+0.93}_{-0.85}& 3.52^{+1.04}_{-0.97}& 2.8^{+0.99}_{-0.97}& 234^{+20}_{-22}& 182^{+45}_{-39} \\
0.0 & 7.77^{+1.0}_{-0.95}& 5.81^{+0.78}_{-0.76}& 3.48^{+0.94}_{-0.92}& 2.76^{+0.96}_{-0.93}& 219^{+17}_{-18}& 173^{+23}_{-21} \\
-0.05 & 7.48^{+1.0}_{-0.97}& 7.02^{+0.93}_{-0.89}& 3.82^{+1.17}_{-1.14}& 3.08^{+1.06}_{-1.05}& 183^{+29}_{-28}& 230^{+28}_{-24} \\
-0.1 & 5.53^{+1.17}_{-1.23}& 6.85^{+1.43}_{-1.24}& 4.38^{+1.67}_{-1.46}& 2.46^{+1.47}_{-1.43}& 288^{+82}_{-67}& 331^{+86}_{-62} \\
-0.2 & 4.22^{+0.88}_{-0.9}& 6.01^{+0.96}_{-0.84}& 3.09^{+1.35}_{-1.26}& 2.83^{+1.23}_{-1.24}& 283^{+95}_{-85}& 410^{+89}_{-66} \\
\hline
\multicolumn{7}{c}{Position 3} \\
\hline
0.15 & 3.4^{+0.59}_{-0.58}& 3.8^{+0.56}_{-0.55}& 2.39^{+0.66}_{-0.64}& 2.67^{+0.66}_{-0.64}& 69^{+38}_{-36}& 249^{+32}_{-25} \\
0.05 & 6.85^{+0.96}_{-0.96}& 8.22^{+0.94}_{-0.91}& 4.4^{+1.08}_{-1.06}& 5.73^{+1.14}_{-1.11}& 49^{+28}_{-28}& 245^{+26}_{-22} \\
0.0 & 9.31^{+1.02}_{-1.01}& 9.86^{+0.91}_{-0.89}& 5.09^{+0.99}_{-0.98}& 6.38^{+1.06}_{-1.05}& 50^{+19}_{-19}& 221^{+17}_{-16} \\
-0.05 & 8.96^{+1.07}_{-1.05}& 9.71^{+0.94}_{-0.93}& 4.56^{+1.18}_{-1.17}& 5.43^{+1.17}_{-1.17}& 87^{+27}_{-27}& 259^{+24}_{-21} \\
-0.1 & 8.46^{+1.11}_{-1.09}& 8.66^{+1.05}_{-0.99}& 4.51^{+1.31}_{-1.25}& 3.5^{+1.27}_{-1.29}& 187^{+37}_{-38}& 289^{+43}_{-32} \\
-0.175 & 4.3^{+0.7}_{-0.69}& 3.72^{+0.6}_{-0.59}& 3.34^{+0.81}_{-0.8}& 2.05^{+0.72}_{-0.7}& 238^{+27}_{-26}& 200^{+27}_{-24} \\
-0.275 & 3.42^{+0.64}_{-0.61}& 2.75^{+0.58}_{-0.55}& 1.98^{+0.63}_{-0.6}& 1.01^{+0.63}_{-0.56}& 149^{+24}_{-23}& 157^{+28}_{-26} \\
\enddata
\tablecomments{The fluxes of [\nii{}] emission lines were fixed at a 3:1 ratio. The Gaussians used to fit the emission lines were assigned to have the same velocity and velocity dispersion (\S~\ref{sec:ana}).
The 1$\sigma$ uncertainties reported in the table were calculated from the parameter posterior distributions (Figure~\ref{fig:corner_central}).}
\end{deluxetable*}


We extracted 1D spectra from the continuum subtracted 2D STIS image, in which the emission lines of interest are detected. From each slit location, we were able to extract spectra out to $\sim0.3\arcsec$ (or $\sim570$ pc), beyond which no signal was detected with STIS. The spectra from the outer pixels have also been combined (up to three) to improve the signal-to-noise ratio. We modeled the emission lines with the Gaussians using the \texttt{Sherpa} Python package.

The emission lines were simultaneously fit using five Gaussian functions, with each function representing one of the following emission lines: \nii{}$\lambda$6548, \nii{}$\lambda$6584, H$_\alpha\lambda$6563, \sii{}$\lambda$6716, and \sii{}$\lambda$6731. During the fitting process, the following constraints were applied: (1) the velocity and velocity dispersion were kept consistent across all individual emission lines, and (2) the flux ratio between the \nii{}$\lambda\lambda6548,6584$ lines was constrained to 3:1, according to theoretical predictions. The first is because all these lines trace the same gas temperature of $\sim10^4$K, and should therefore be subject to the same kinematics. 

We also tested alternative model configurations where the kinematic properties of the emission lines were allowed to vary freely. Specifically, we let the velocity and velocity dispersion parameters vary between individual lines. However, this did not significantly improve the fit. We also tested a model that included two Gaussian components for each emission line, but this approach similarly did not produce significantly better results for our STIS data.
More specifically, we tested fitting a broad line in addition to the narrow lines of each emission lines, but this did not give a good fit to the spectra.
Consequently, the data are consistent with a single Gaussian for each emission line.
We did not include skewness as a fit parameter for the Gaussian profile, as the data did not have adequate SNR to robustly constrain this parameter.
Finally, we used \texttt{emcee} to perform posterior sampling, allowing us to extract posterior distributions of the fit parameters.
To ensure MCMC convergence, we ran a total of 10000 steps with 200 chains (\texttt{emcee} walkers) and discarded the first 5000 steps as burn-in iterations for each extracted spectrum.
We inspected the chains to verify that the posterior probabilities were well-converged.
Furthermore, convergence was confirmed using the $\hat{R} \approx 1$ criterion \citep{Gelman1992}, indicating that the between-chain and within-chain variances are approximately equal.
From these distributions, we obtained the best-fitting values and associated uncertainties. 

The spectra and individual fits for all pixels with detected emission lines are shown in Fig. \ref{fig:best_fit_all}. In Fig. \ref{fig:corner_central}, we present the posterior distributions of the fit parameters for the central pixel of the central slit (Position 1), which is centered on the nucleus, as an example of the fits. Additionally, Table~\ref{tab:fit_table} displays the emission line fluxes and kinematics with uncertainties for each pixel with detections.


\begin{figure*}
\centering
\includegraphics[width = 0.495\textwidth]{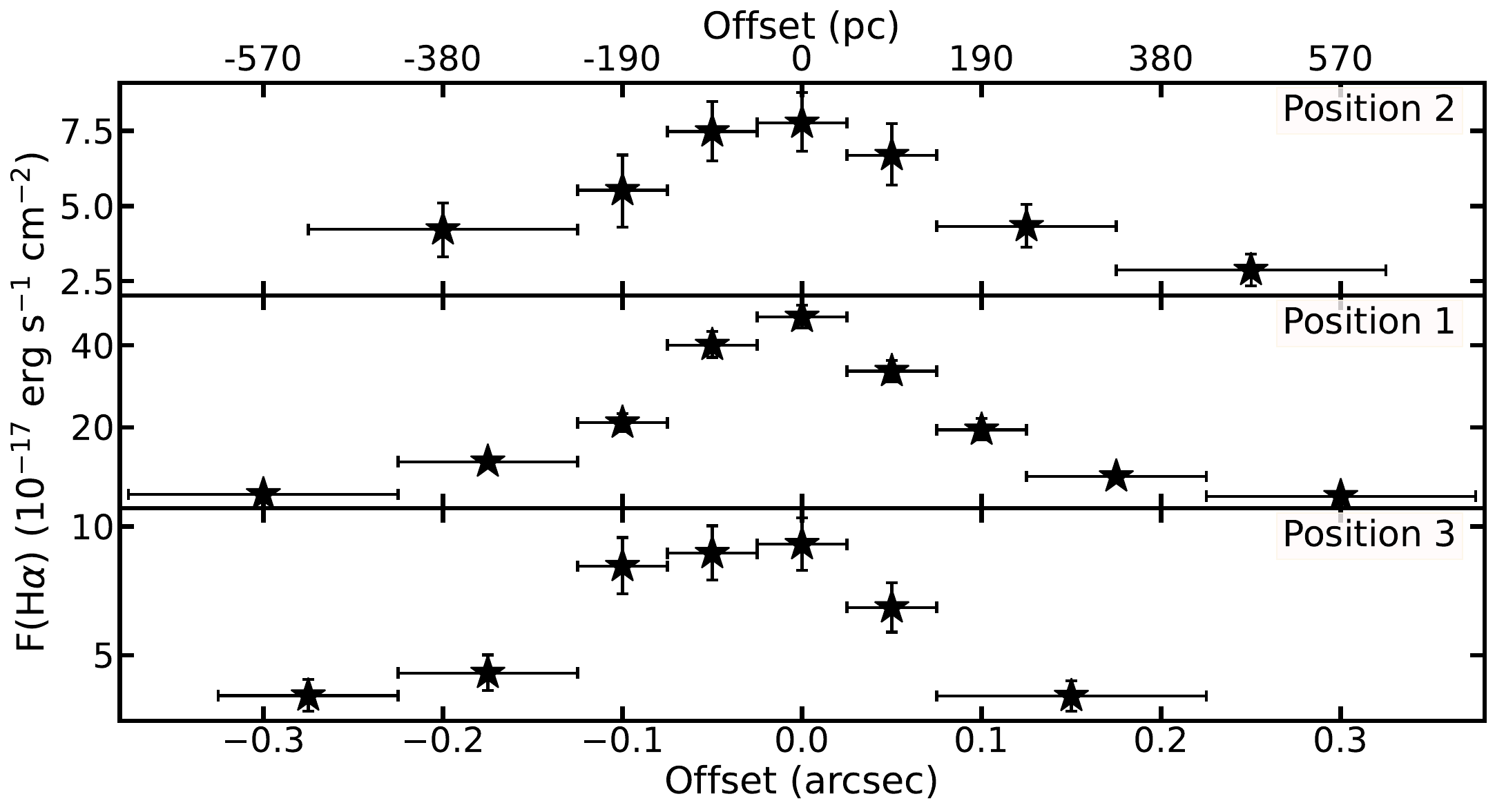}
\includegraphics[width = 0.495\textwidth]{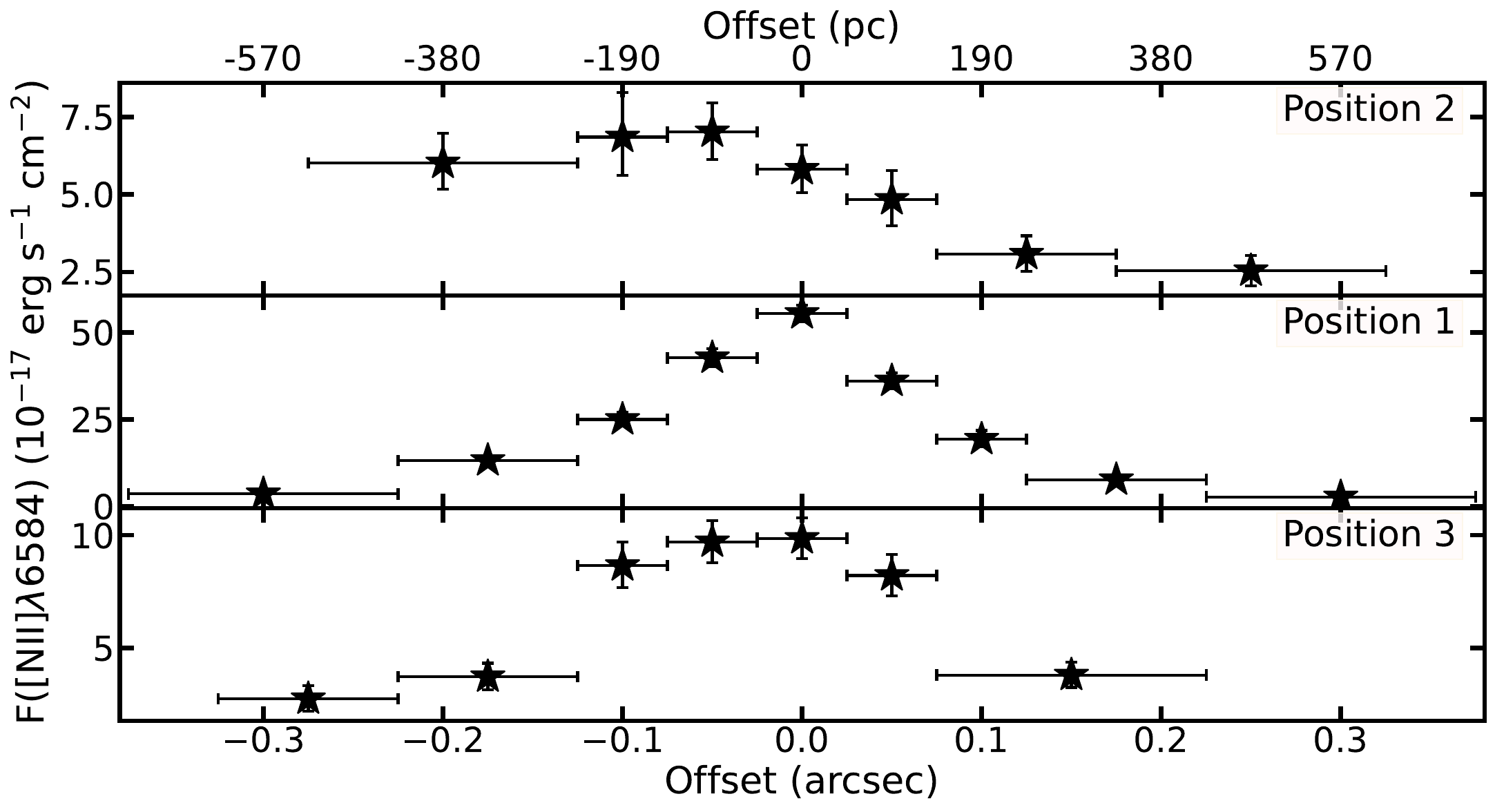}
\includegraphics[width = 0.495\textwidth]{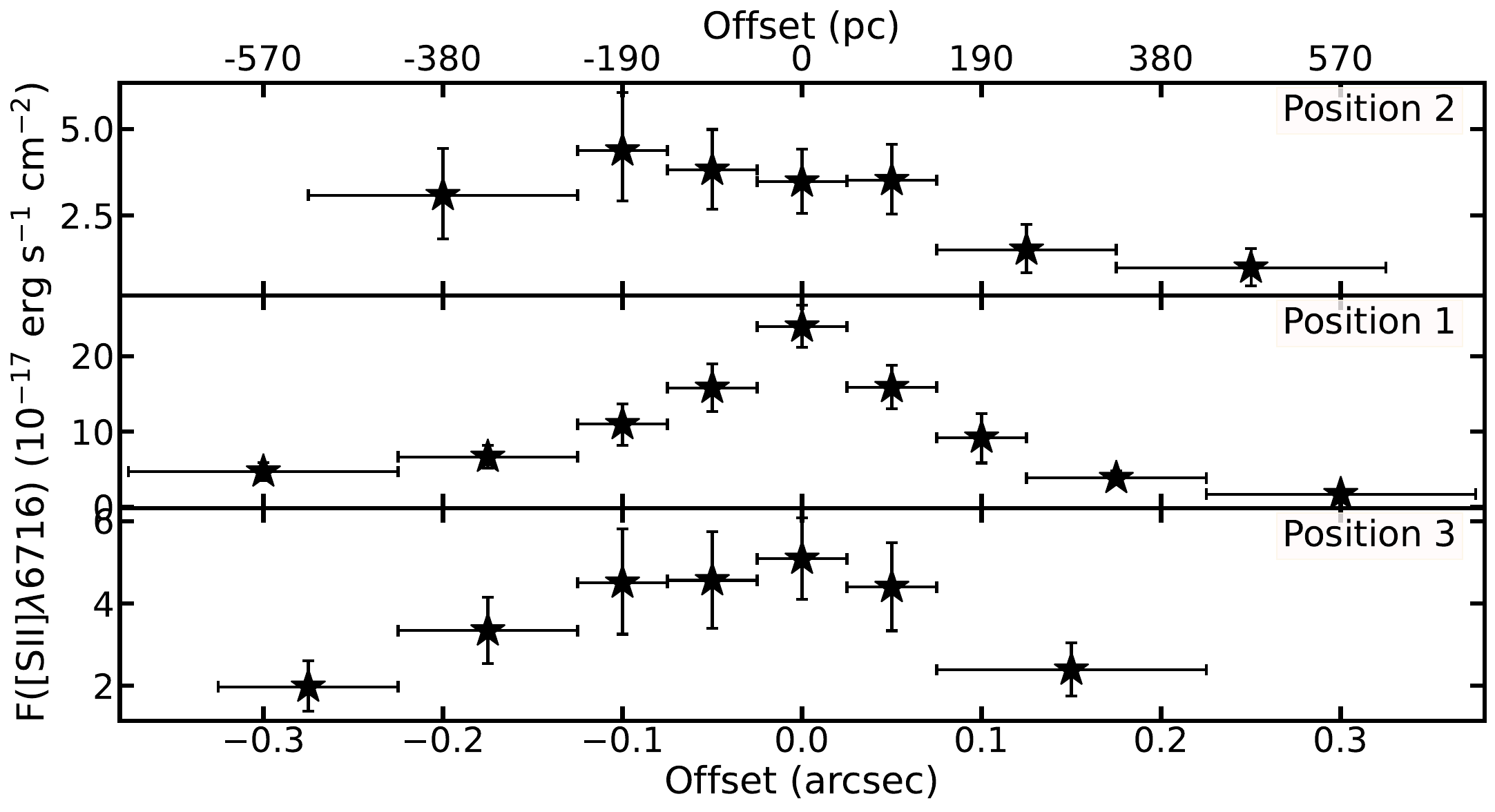}
\includegraphics[width = 0.495\textwidth]{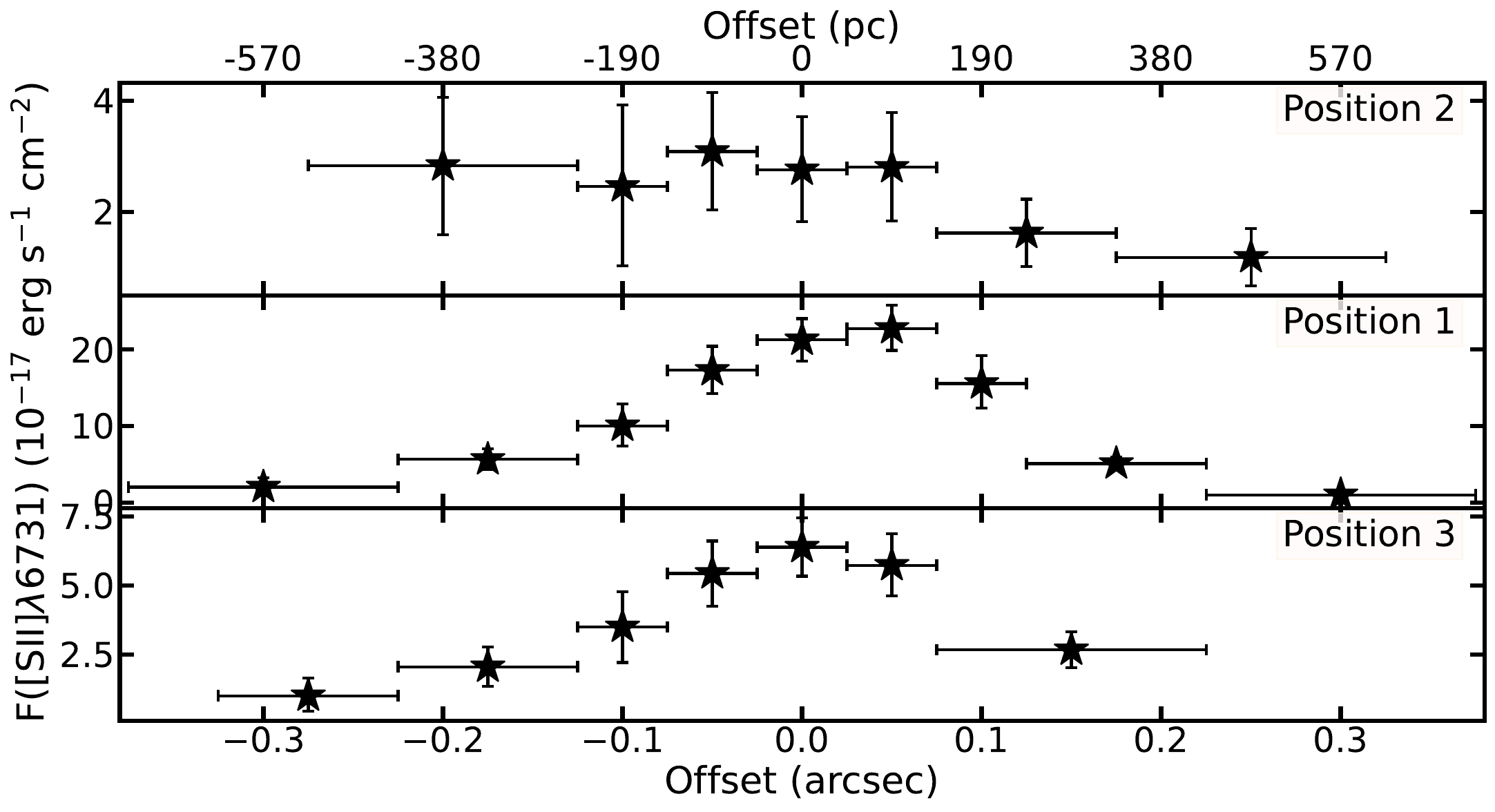}
\caption{The flux measured from
H$_\alpha\lambda$6563, \nii{}$\lambda$6584, \sii{}$\lambda$6716, and \sii{}$\lambda$6731
emission lines as a function of location along the three slit positions.
The vertical error bars represent the $1\sigma$ measurement uncertainties, and the widths of the pixels from which the spectra were extracted are shown as horizontal error bars.
In the spectral models, the flux ratio between the [\ion{N}{2}]$\lambda\lambda6548,6584$ was held at 1:3.}
\label{fig:em_flux}
\end{figure*}

\section{Results} \label{sec:res}

\begin{figure}
\centering
\includegraphics[width = 1\columnwidth]{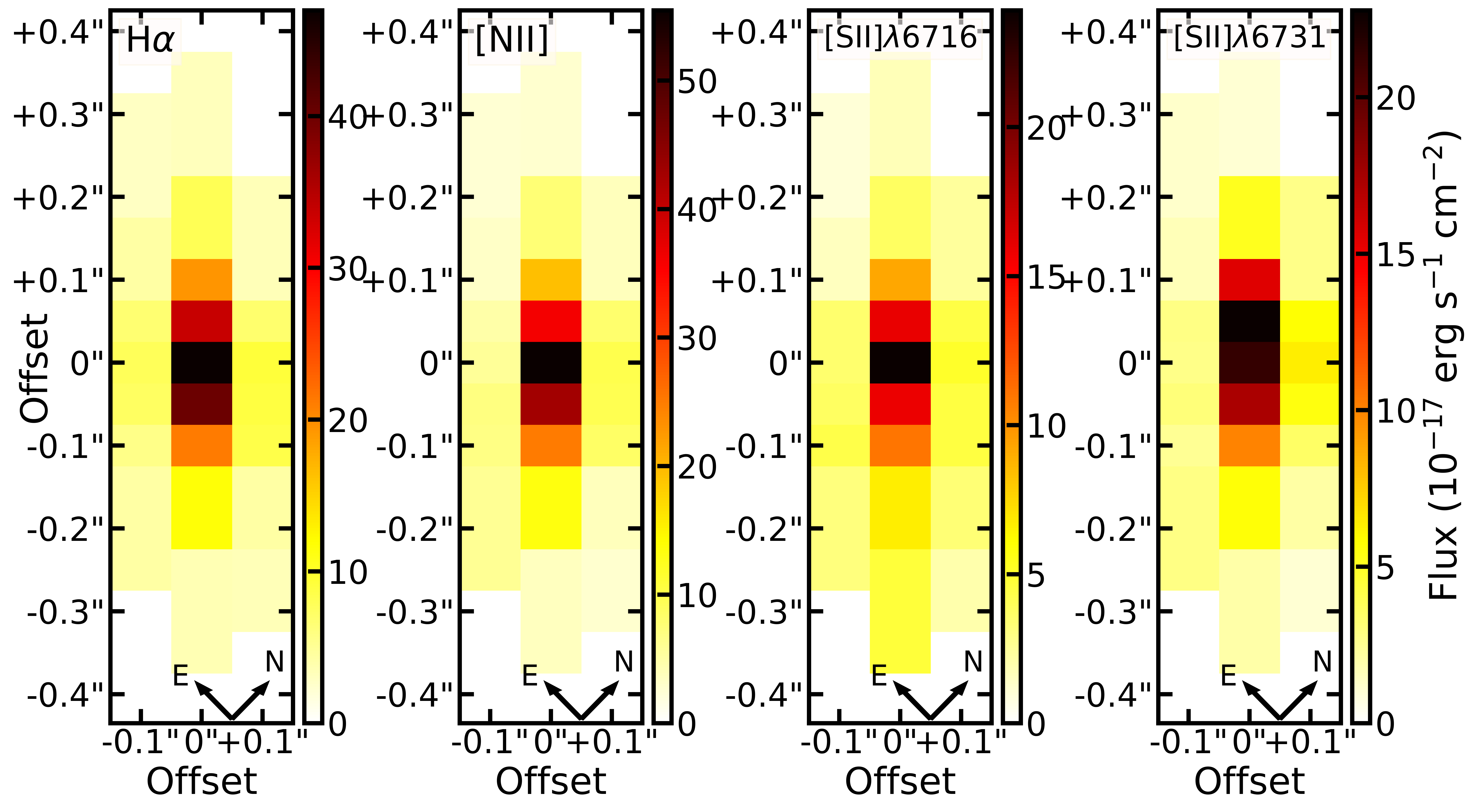}
\caption{Two-dimensional flux maps for the emission lines detected in \textit{HST}/STIS observation of \zn.
The highest emission line fluxes are observed at the central spexels and generally decrease away from the center.
The N-E coordinate directions are represented in the figure that reflects the position angle used with the STIS slits}
\label{fig:3d_flux_maps}
\end{figure}

\subsection{Emission lines detected} \label{sec:reslines}

In Fig. \ref{fig:em_flux}, we present the radial profiles of the emission line fluxes for \nii{}$\lambda$6584, H$_\alpha\lambda$6563, \sii{}$\lambda$6716, and \sii{}$\lambda$6731. The flux ratio between the [\ion{N}{2}]$\lambda\lambda6548,6584$ lines is fixed to 3:1, so we only show the results for one line. Additionally, Fig. \ref{fig:3d_flux_maps} displays the 2D flux maps for the same lines.

Both figures clearly demonstrate that the flux peaks at the center, corresponding to the location of the nucleus. In all emission lines, there is a steady and consistent increase in flux towards the nucleus, increasing by a factor of 10 within the central 0.3 \arcsec (or 570 pc). The nucleus or AGN location, therefore, contains significantly brighter ionized gas, as traced by \nii{}$\lambda$6548, H$_\alpha\lambda$6563, \sii{}$\lambda$6716, and \sii{}$\lambda$6731. Beyond approximately 0.3 \arcsec (or 570 pc), there is no detection of emission from our STIS data, although the HST FR716N image does show the presence of diffuse ionized gas that extends beyond 0.3 \arcsec along our slit positions.

A factor of 10 decrease is also observed in the HST FR716N image shown in Fig. \ref{fig:zoom}. The H$\alpha$ emission drops significantly beyond a radius of 0.5 \arcsec (or 950 pc) in the South-West, and particularly in the North-East direction. In contrast, the emission is more extended in the South-East to North-West direction. At larger scales, beyond a radius of 0.5 \arcsec (or 950 pc), as shown in Fig. \ref{fig:mediumscale}, the H$\alpha$ emission appears to trail behind the X-ray cavities, which are located in the South-East and North-West directions.

We also note that the strengths of the \nii{}$\lambda$6584 line appear comparable or greater to those of the H$_\alpha\lambda$6563 emission line across the pixels in Position 1 that encompasses the nucleus. More details on this can be found in \S~\ref{sec:reslineratios} and \S~\ref{sec:disc_flow_ion}.


\subsection{Kinematics} \label{sec:reskin}

\begin{figure*}
\centering
\includegraphics[width = 0.495\textwidth]{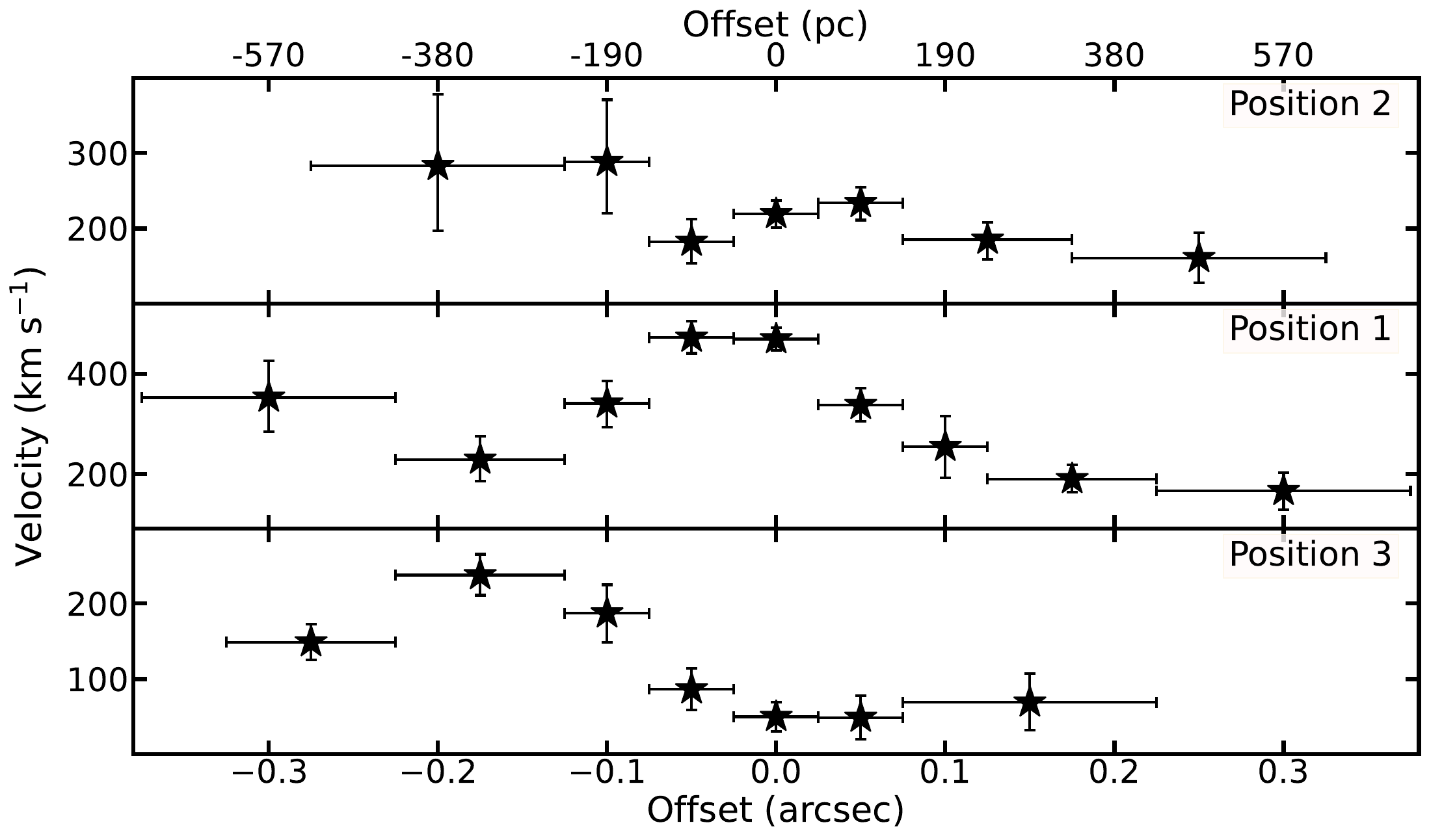}
\includegraphics[width = 0.495\textwidth]{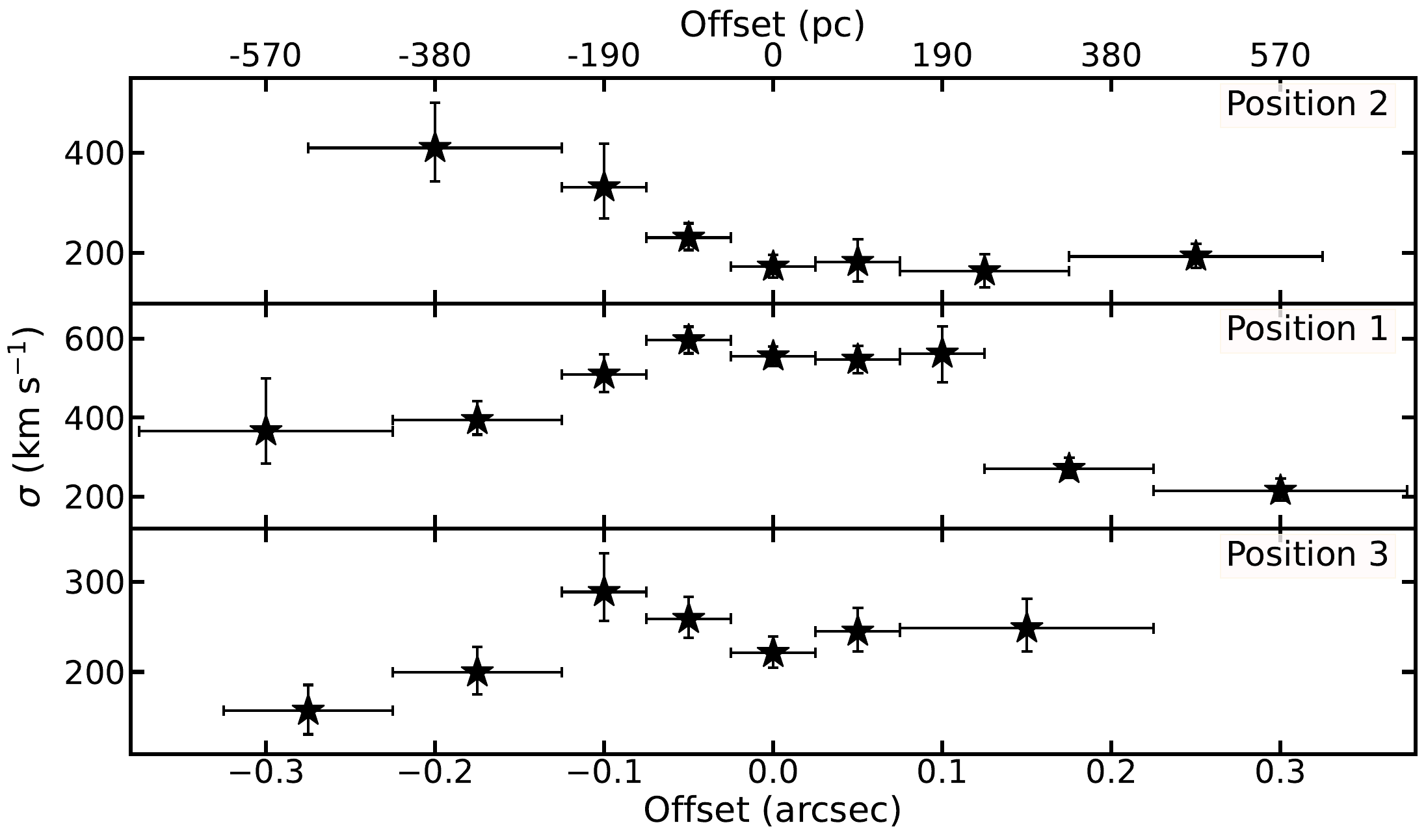}
\caption{The velocity and velocity dispersion measured from the emission lines as a function of pixel location along the three slit positions. We note that our spectral model required the kinematic parameters for all the detected emission lines to be the same.
The highest velocity and velocity dispersion values are found at the central pixel in the vicinity of the central AGN. The velocity of the emission lines are all positive with respect to the systematic redshift of the BCG. The error bars are as in Figure~\ref{fig:em_flux}.}
\label{fig:vel}
\end{figure*}

In Fig. \ref{fig:vel}, we present the velocity and velocity dispersion for each pixel that has a detection as a function of radius. The first observation is that all velocities are positive, with an overall mean of $\sim225$ km s$^{-1}$.
This is with respect to a reference redshift of $z=0.102428$, determined from the systemic redshift of the galaxy, calculated using the stellar velocity in the inner-most region of the galaxy based on stellar absorption lines and stellar population synthesis model fitting \citep{Marie-Joelle}.
They focused specifically on the goodness of fit for the Ca H \& K lines, which are primarily from old stellar populations that are expected to be strongly gravitationally bound to the central galaxy.
We note that the velocity measurements could be lowered by $\sim120\ \mathrm{km\ s^{-1}}$ if a systemic redshift of $z=0.1028$ were adopted, which is a more commonly used value in previous studies of \zn\/ \citep{Russell2016,Hunstead1978,Olivares2019}.
This would also make the line emissions in some of the STIS slits (mainly Position 3) appear blueshifted with negative velocities.

\begin{figure}
\centering
\includegraphics[width = 1\columnwidth]{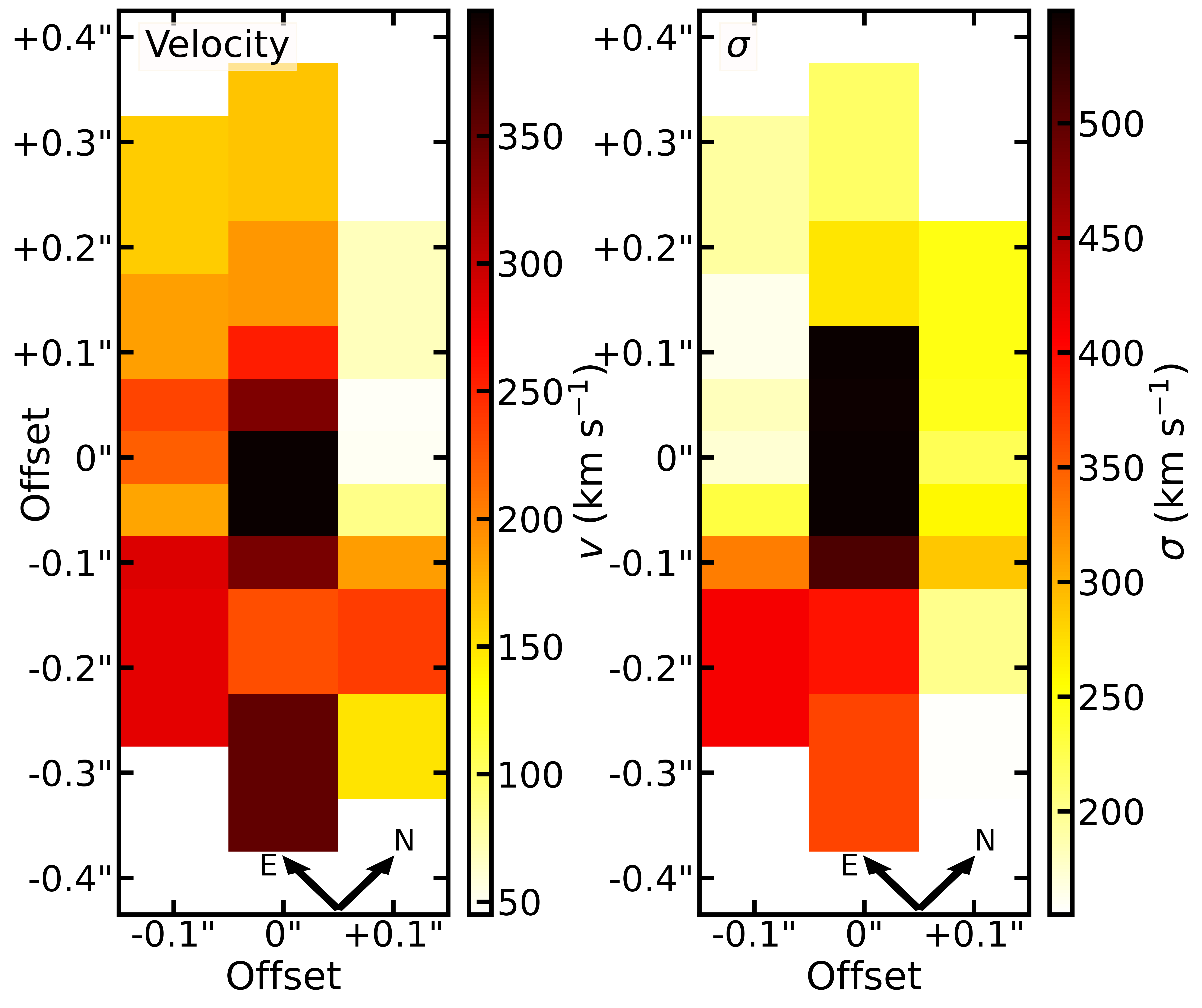}
\caption{Two-dimensional maps of the kinematic properties.
The central spexels covering the core of the BCG show the highest values of velocity and velocity dispersion.}
\label{fig:3d_kin_maps}
\end{figure}

On average, Positions 1, 2, and 3 have mean velocities of $\sim312$ km s$^{-1}$, $\sim220$ km s$^{-1}$, and $\sim119$ km s$^{-1}$, respectively. The largest shift is observed on the South-West side of the nucleus in the negative offset of Position 1. Hence, the ionized gas on scales of hundreds of parsecs appears to be highly redshifted by several hundred km s$^{-1}$ compared to the average velocity of the ionized gas in the galaxy measured on kpc scales. The highest offset is associated with the pixel at the location of the nucleus (central pixel in Position 1), with a redshifted velocity of $\sim470$ km s$^{-1}$. This is further highlighted in Fig. \ref{fig:3d_kin_maps}.
We discuss this further in \S~\ref{sec:disc_flow_kin}.

The slit positioned along the nucleus (Position 1) also shows a large increase of $\sim300$ km s$^{-1}$ in velocity in both the South-West and North-East directions up to the nucleus within the inner 0.3\arcsec (or 570 pc). We further discuss this increase and its implications in terms of free-fall gas in \S~\ref{sec:disc_flow_kin} and \S~\ref{sec:disc_obs}.

In terms of velocity dispersion, as shown in Fig. \ref{fig:vel} and Fig. \ref{fig:3d_kin_maps}, there is a similar trend in the sense that the velocity dispersion increases strongly towards the nucleus. Indeed, as seen along Position 1, the velocity dispersion increases by almost 400 km s$^{-1}$ within 0.3\arcsec (or 570 pc). The peak is reached at the location of the nucleus and implies a large FWHM of $\sim1400$ km s$^{-1}$, consistent with the location of a very massive supermassive black hole (which we discuss in \S~\ref{sec:disc_BHmass}). On average, Positions 1, 2, and 3 have mean velocity dispersions of $\sim445$ km s$^{-1}$, $\sim241$ km s$^{-1}$, and $\sim232$ km s$^{-1}$, respectively. Overall, the average velocity dispersion across all pixels is $\sim320$ km s$^{-1}$.


\subsection{Line ratios} \label{sec:reslineratios}

\begin{figure}
\centering
\includegraphics[width = 1\columnwidth]{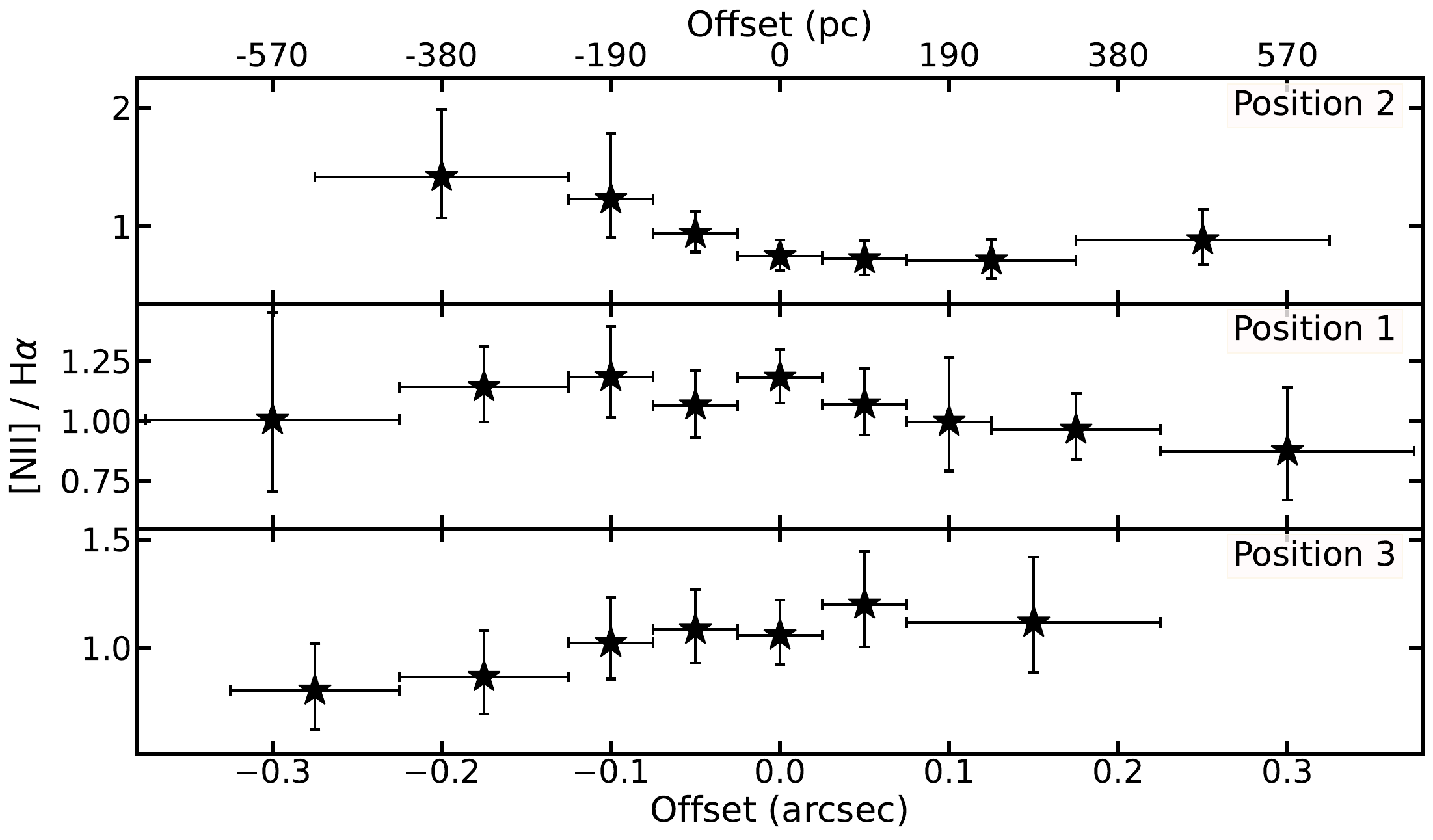}
\caption{[\ion{N}{2}]$\lambda6584$/H$\alpha$ emission line intensity ratio as a function of location along the three slit positions. The ratio remains above unity across the pixels, which indicates that the ISM responsible for producing the observed emission lines is illuminated with hard SED, likely originating from the AGN in the BCG.
The error bars are as in Figure~\ref{fig:em_flux}.}
\label{fig:nhrat_flux}
\end{figure}
\begin{figure}
\centering
\includegraphics[width = 1\columnwidth]{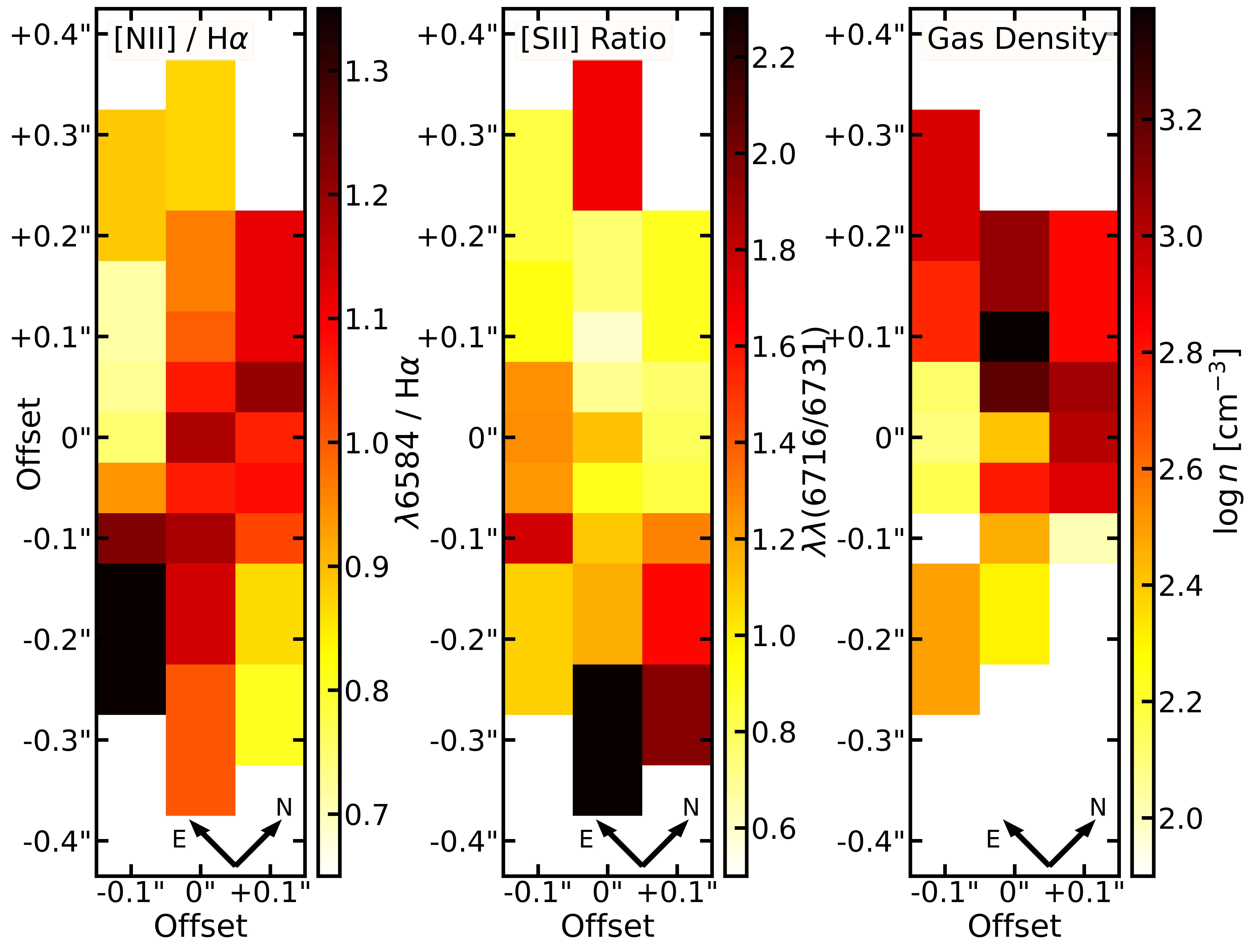}
\caption{Two-dimensional maps of [\ion{N}{2}]$\lambda6584$/H$\alpha$ and [\ion{S}{2}]$\lambda\lambda6716, 6731$ doublet line ratios, and ISM density estimated from the [\ion{S}{2}] line ratios.
We find a modest indication for higher [\ion{N}{2}]/H$\alpha$ ratios along the north-south direction across the central BCG and for density increase towards the northeast. However, these trends are observed within the uncertainties.}
\label{fig:3d_ratio_maps}
\end{figure}

In Fig. \ref{fig:nhrat_flux}, we show the ratio between the intensity of the [\ion{N}{2}]$\lambda6584$ and the intensity of H$\alpha$ emission line for each of the three slit positions. The ratio between [\ion{N}{2}]$\lambda6584$ and H$\alpha$ is often used as a proxy for estimating the hardness of the SED of the illuminating source \citep[e.g.,][]{Baldwin1981,SanchezAlmeida2012}. We did not find a clear peak in the distribution of the ratios at the center near where the BCG is located. In Fig. \ref{fig:3d_ratio_maps}, we also highlight in the left panel the ratio map. Essentially, the ratio hovers around unity throughout the core region of the BCG. Positions 2 and 3 in Fig. \ref{fig:nhrat_flux} seem to suggest that there is a gradient along the slits, but the values are within 2$\sigma$ of each other. Nevertheless, the high ratios we obtained throughout the pixels ([\ion{N}{2}]/H$\alpha\gtrsim1.0$) suggest that the ISM near the vicinity of the BCG is excited by the central AGN or another hard source \citep[e.g.,][]{Freitas2018}.

\begin{figure*}
\centering
\includegraphics[width = 0.495\textwidth]{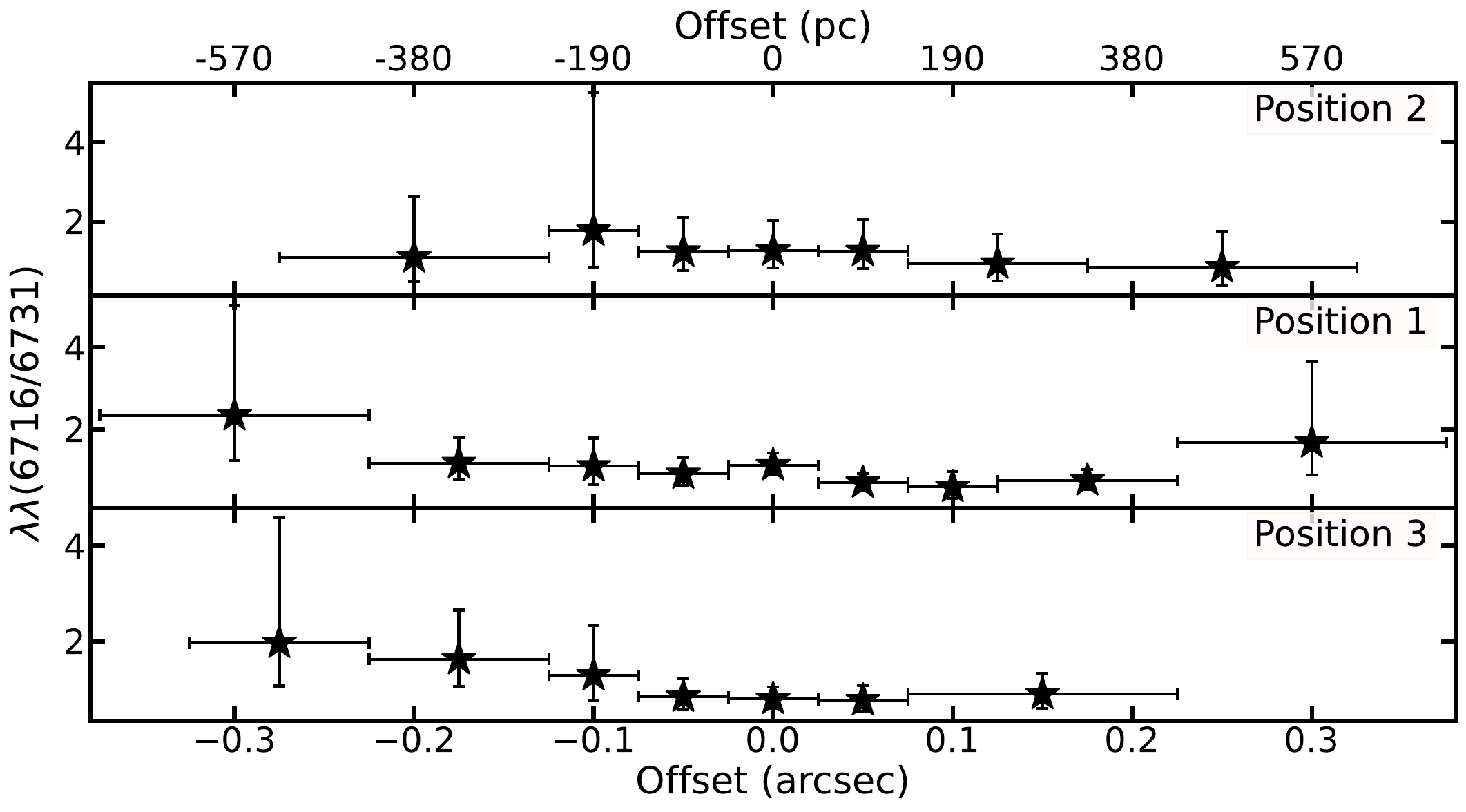}
\includegraphics[width = 0.495\textwidth]{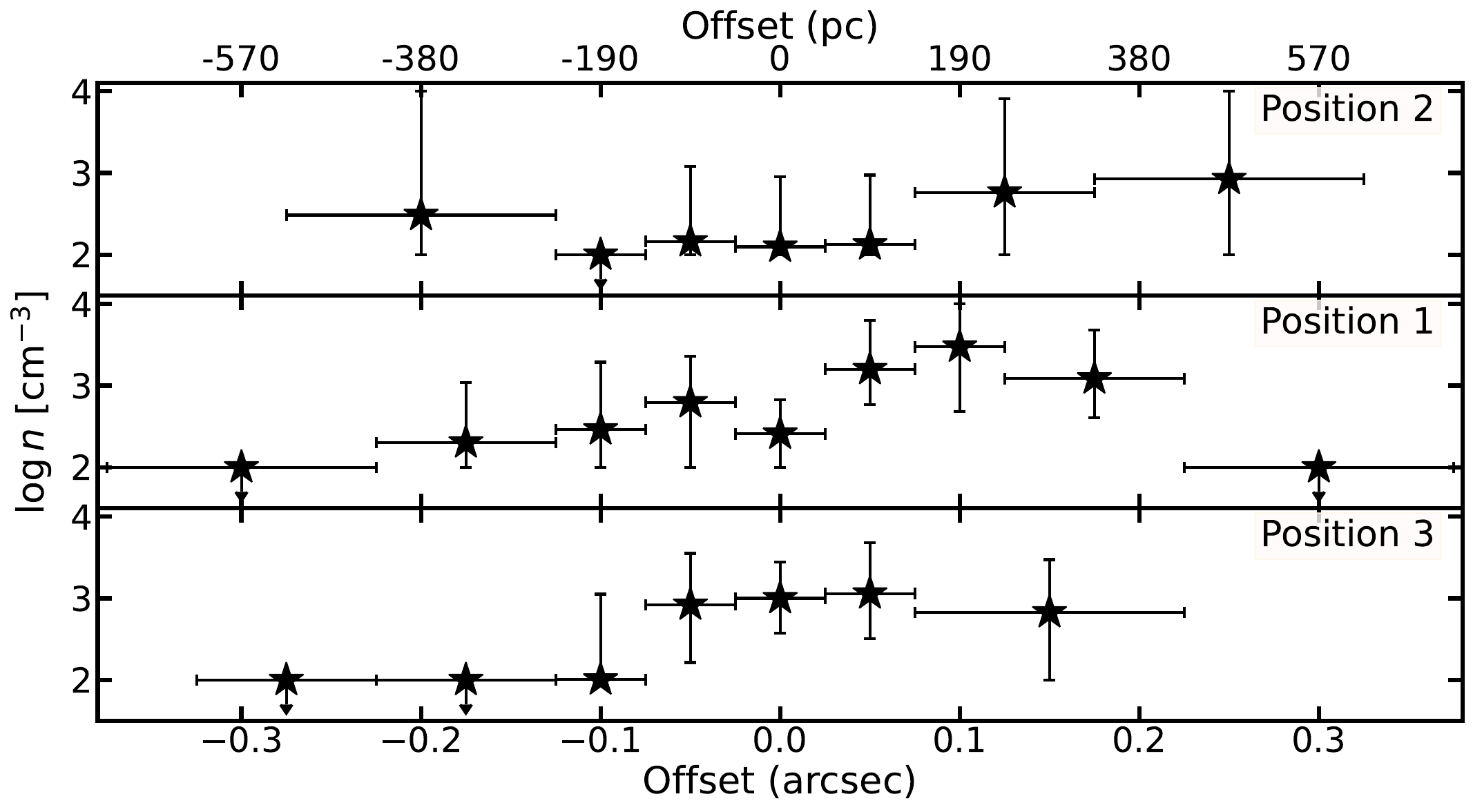}
\caption{[\ion{S}{2}]$\lambda\lambda6716, 6731$ doublet line ratio and the corresponding ISM density as a function of location along the three slit positions.
The line ratios and corresponding densities show near-flat distributions within the errors with a slight enhancement in density towards positive offset positions.
The error bars are as in Figure~\ref{fig:em_flux}.}
\label{fig:s2}
\end{figure*}

In the left panel of Fig. \ref{fig:s2}, we show the \sii{}$\lambda\lambda6716, 6731$ doublet line ratio along the three slit locations. In the right panel of the same figure, we show the corresponding ISM density, which was estimated by comparing the observed \sii{} ratio with that calculated from photo-ionization simulations (\texttt{Cloudy}; \citealt{Ferland2013}) assuming an ISM temperature of T=$10^{4}$ K \citep[e.g.,][]{Revalski2018,Revalski2022}. We also show these values in the middle and right panels of Fig. \ref{fig:3d_ratio_maps}. Essentially, these figures reveal that the density is remarkably uniform in the core of \zn\/, despite the presence of an actively accreting AGN in the center of the BCG.
We further discuss this and its implications for accretion physics in \S~\ref{sec:disc_flow_den}.



\section{Discussion} \label{sec:disc}

This study presents one of the first detailed observations of gas dynamics within the core of a BCG with powerful AGN feedback. Below, we offer an in-depth analysis of several key results derived from these observations, highlighting their implications and significance in the context of the physics of AGN feedback.

\subsection{Black hole mass estimate}\label{sec:disc_BHmass}

In Section \ref{sec:disc_flow_kin}, we explore the highly chaotic nature of the ionized gas flow. The lack of organized motion, such as rotation, makes it impossible to obtain a robust dynamical mass measurement of the SMBH. Instead, we attempt a preliminary estimate of the SMBH mass ($M_{\rm BH}$) using the velocity dispersion map shown in Fig. \ref{fig:3d_kin_maps}. The velocity dispersion peaks near the AGN's location, reaching a maximum value of $\sigma=595\pm32$ km s$^{-1}$, which corresponds to a FWHM of $\sim1400$ km s$^{-1}$. Assuming that the virial theorem holds—though this is uncertain given the chaotic conditions—the corresponding $M_{\rm BH}$ can be approximated as $\sim\sigma^2R/G$, where $R$ is the radius within which $\sigma$ is measured, i.e., the width of the slit $\sim0.1''$ (or 188 pc), and $G$ is the gravitational constant. Using $\sigma\sim595$ km s$^{-1}$, this yields an estimated (and very approximate) $M_{\rm BH}$ of $\sim1.5\times10^{10}M_\odot$. The sphere of influence for a SMBH is given by $r_{\rm inf} = GM_{\rm BH}/\sigma^2$. For an SMBH mass of $\sim1.5\times10^{10}M_\odot$, the sphere of influence is estimated to be $r_{\rm inf} \sim 0.4\arcsec$ (or 800 pc) for our target. This suggests that the sphere of influence is resolved and covers roughly a dozen STIS pixels, and as illustrated in Fig. \ref{fig:zoom}, the kinematics appear resolved within this region. However, given the chaotic nature of the dynamics, it remains unclear whether the virial theorem provides a reliable mass estimate in this context.

There are however several other additional factors that lead us to conclude that \zn\/ likely hosts a very massive $\sim10^{10}M_\odot$ SMBH. First, \zn\/ is located in one of the most massive galaxy clusters known. Since BCG mass/luminosity, and consequently SMBH mass, scales with cluster mass \citep[][]{Stott2012}, BCGs in the most massive clusters are expected to host some of the most massive SMBHs \citep[][]{Natarajan2009}. Notably, some of the most massive SMBHs known to date reside in similar systems such as NGC 4889 \citep[$M_{\rm BH} = 2.1 \times 10^{10}M_\odot$;][]{McConnell2012}, the BCG of the massive Coma cluster of galaxies. \zn\/ is also located in one of the strongest cool core galaxy clusters known, where the BCG drives powerful mechanical AGN feedback with $P_{\rm cav}\sim5\times10^{45}$ erg s$^{-1}$ \citep[][]{Rafferty2006}. To power the outbursts over cosmic time, the central SMBH must have accreted far more than $10^{9}M_\odot$, allowing us to place a lower limit on its mass \citep[][]{JHL2012a}.

Second, in \citet{JHL2012a}, we explored the location of BCGs on the fundamental plane of black hole activity, a well-established relation between SMBH mass, radio luminosity and X-ray luminosity \citep{Merloni2003}. By measuring the nuclear radio and X-ray luminosities of BCGs, we estimated their black hole masses. We found that the masses derived from the standard $M_{\rm BH}-\sigma$ relation were systematically underestimated by a factor of 10, on average. Crucially, we predicted that BCGs in massive, strong cool core clusters, such as \zn, require $10^{10-11}M_\odot$ black holes to be consistent with the fundamental plane. For \zn, this factor of 10 increase suggests a black hole mass of $\gtrsim10^{10}M_\odot$. Confirming that massive cool core clusters host such massive black holes would greatly impact our understanding of galaxy evolution in clusters. Currently, very few BCGs in strong cool core clusters have dynamical SMBH mass measurements. For instance, \citet{DallaBonta2009} measured the SMBH masses of three BCGs, though these were not located in strong cool core clusters. In contrast, \citet{Mehrgan2019} provided an example of an ultra-massive $10^{10}M_\odot$ SMBH in a strong cool core cluster. Note also, the existence of the ultra-massive $3.27\pm2.12\times10^{10}M_\odot$ SMBH in Abell 1201 \citep[][]{Nightingale2023}, although not a cool core cluster.

Finally, we analysed the stellar surface brightness profile of \zn\/ from the HST F814W image (shown in Fig. \ref{fig:mediumscale}), finding it to be dominated by one of the largest cores known. Cores, or light deficits, appear as breaks in the stellar surface brightness profile. The leading theory suggests they form from the decay of binary SMBHs, which in turn expel stars from the galaxy \citep[e.g. ][]{Beg1989,Natarajan2009}. Importantly, core sizes are known to correlate with the total mass of stars ejected (light deficit), and thus, the mass of the central SMBH \citep[e.g. ][]{Lauer2007}. The size of the core in \zn, derived by fitting standard Nuker profiles to the surface brightness, is among the largest in any BCG: $\sim3$ kpc. A rough estimate of the corresponding SMBH mass, based on this light deficit, also yields $5\times10^{10}M_\odot$. 

Together, these multiple pieces of evidence indicate that \zn\/ may host one of the most massive SMBHs in the Universe. Notably, for a SMBH with a mass of $\sim10^{10}M_\odot$, the shadow would be approximately 3 micro-\arcsec. This is beyond the reach of the current Event Horizon Telescope \citep[EHT; e.g.][]{2024EHT}, but it could be within the reach of the Black Hole Explorer (BHEX), a new mission under development that aims to extend the EHT to space and is expected to launch within the next decade \citep[e.g.,][]{Johnson2024}. The BHEX could probe scales as small as $4-5$ micro-\arcsec, suggesting that it may potentially resolve the shadow in \zn\/ or, at the very least, the jet-launching region. The next-generation Event Horizon Telescope (ngEHT), with an angular resolution of $\sim10-15$ micro-\arcsec \citep[e.g.][]{Lico2023}, could also resolve features on the scale of the SMBH's shadow in \zn, or at least the disk-jet structure. These telescopes offer a rare and unique opportunity to probe AGN feedback in radio-mode down to its very core. Note that the radio core in \zn\/ \citep[$\sim$5 mJy at 150 GHz][]{Hogan2015b} is at the limit of observability for ngEHT and BHEX, but that there are many other BCGs in cool core clusters that could serve as excellent targets for this telescope whether to image the shadow or even search for absorbing clouds along the line of sight that appear to be feeding the SMBH \citep[][]{Rose2019a,Rose2019b}.  

It is important to note, however, that in the absence of a direct dynamical measurement of the black hole’s mass, our estimate of $1.5 \times 10^{10} M_\odot$ remains uncertain and should be regarded as preliminary. Further observational efforts will be necessary to refine this measurement and confirm the exceptional nature of the central black hole in \zn.

\subsection{Accretion flow: kinematics}\label{sec:disc_flow_kin}

In \S~\ref{sec:reskin}, we presented the kinematical maps of the ionized gas as traced by \nii{}$\lambda$6548, \nii{}$\lambda$6584, H$_\alpha\lambda$6563, \sii{}$\lambda$6716, and \sii{}$\lambda$6731. These emission lines primarily trace ionized gas at $\sim10,000$ K, probing only one phase of the multiphase gas in the BCG.

Our observations clearly indicate that the flow of this gas phase is highly chaotic and disturbed on the sub-kpc scales probed by STIS, extending all the way to the central pixel (width of the slit is $\sim188$ pc). In \S~\ref{sec:disc_obs}, we discuss how this flow connects to the kinematics of the larger kpc scales. Here, we focus only on the STIS results.

\begin{figure*}
\centering
\includegraphics[width = .7\textwidth]{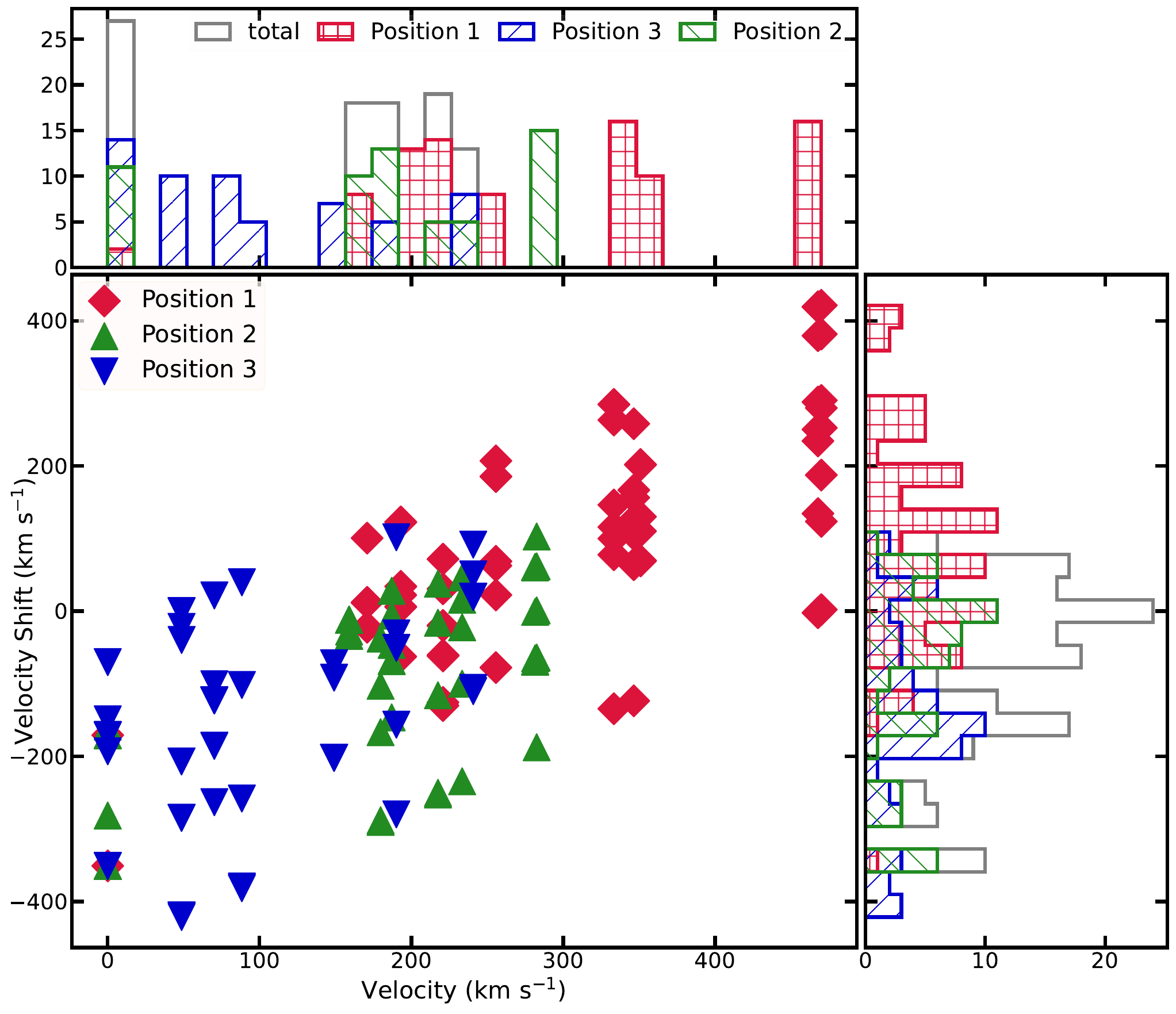}
\caption{Distribution of velocity shifts between neighboring STIS pixels as a function of velocity.
The chaotic nature of ionized gas kinematics is highlighted with velocity shifts reaching up to $\pm 400\ \mathrm{km\ s^{-1}}$ between adjacent pixels.
Position 1, which encompasses the nucleus, shows predominantly positive velocity shifts, while Positions 2 and 3 exhibit more negative shifts, highlighting the disturbed flow of gas near the AGN.}
\label{fig:veldiff}
\end{figure*}

To illustrate the chaotic nature of the flow observed with STIS, Fig. \ref{fig:veldiff} shows the distribution of velocity shifts. For each STIS pixel with a detection, we calculate the velocity difference between its velocity and that of its neighbouring pixels. For example, the pixel located at the AGN in Position 1 has 8 neighbouring pixels, resulting in 8 corresponding velocity difference values. We repeat this procedure for every pixel with a detection. Finally, the resulting velocity shifts are plotted as a function of the velocity in Fig. \ref{fig:veldiff}, which also shows the distribution of both velocity shifts and velocities.


This plot reveals two key findings. First, the velocity shift between neighbouring pixels can be substantial, sometimes reaching up to $\pm$ 400 km s$^{-1}$. The spectral lines were modelled with Gaussian profiles, and uncertainties were rigorously estimated using MCMC methods, ensuring their robustness. With uncertainties ranging from $\sim 20-60$ km s$^{-1}$ and observed shifts of several hundred km s$^{-1}$, these variations are likely real. In a well-ordered system, such as a rotating disk, we would expect smooth velocity gradients and minimal differences between neighbouring pixels. In contrast, the figure illustrates the chaotic nature of the flow, where the velocity can jump by hundreds of km s$^{-1}$ between two pixels, corresponding to a physical scale of 188 pc.

Second, the plot reveals a trend: velocities along Position 1 are generally higher than those in neighbouring pixels, resulting in positive velocity shifts. In contrast, Positions 2 and 3, particularly Position 3, exhibit negative velocity shifts. Notably, all three slits are positioned roughly perpendicular to the direction of the radio jets based on the X-ray cavity positions. Since Position 1 runs through the AGN, it may be experiencing greater disturbances, which could explain these observed differences.

The kinematics of the ionized gas within the black hole's sphere of influence, on scales of 188 pc, are therefore highly chaotic in the direction perpendicular to the radio jet. This chaotic behaviour could stem from the intrinsic nature of black hole accretion in radio-mode feedback, or it may result from the interaction between the radio jets of \zn\/ and the surrounding gas. A combination of both mechanisms is also possible, where the radio jets induce chaotic motion, reduce angular momentum, and facilitate the infall of gas toward the SMBH, generating a self-regulated feedback loop. 

This may be a key result since gas with high angular momentum is normally ejected when it encounters the centrifugal barrier of the SMBH \citep{Narayan2011}. To settle into an accretion flow, the gas must therefore lose angular momentum through processes such as thermal instability, turbulence, stellar winds, or cloud-cloud and cloud-filament interactions. Our findings suggest that the flow manages to reduce its angular momentum, possibly through turbulence induced in the direction perpendicular to the radio jet, potentially allowing accretion to occur, at least down to a scale of approximately 188 pc.

The relationship between black hole spin and jet launching is a also key question in understanding the mechanisms driving energetic outflows. \citet{Blandford1977} demonstrated that higher spins lead to the production of stronger jets, emphasizing the importance of spin in powering relativistic outflows. However, chaotic accretion has been shown to spin down SMBHs over cosmic time \citep[e.g.,][]{Ricarte2024,Berti2008}. Recent simulations by \citet{Bus2019} introduce a new sub-grid model for SMBH spin evolution in the moving-mesh code AREPO. This allows the author to trace the evolution of SMBH spin with greater accuracy. They include different forms of accretion through a Shakura-Sunyaev $\alpha$-disc, chaotic accretion, and SMBH mergers. They find that chaotic accretion is more prevalent in higher-mass SMBHs with $M_{\rm BH}>10^8$ M$_\odot$. This is similarly the case for the simulations of \citet{Sala2024}, where they used a sub-resolution prescription that models the SMBH spin implemented in cosmological hydrodynamical simulations.

Given the chaotic flow observed in \zn\/ and the large mass of its SMBH (see Section \ref{sec:disc_BHmass}), this points to the possibility that the spin of \zn\/ may be low. However, PKS 0745-191 hosts a very powerful radio jet capable of carving out large X-ray cavities (see Fig. \ref{fig:largescale}), with an estimated power of $P_{\rm cav}\sim5\times10^{45}$ erg s$^{-1}$. Assuming a 10 per cent jet efficiency, the corresponding accretion rate would be $\sim0.9$ M$_{\odot}$ yr$^{-1}$, although it could be lower if the jet efficiency is higher \citep[see][]{Techkhovskoy2010}. A high spin might therefore be required to sustain the radio jet, as proposed by \citet{McN2009,McN2011}, who argue that high spins are likely necessary to power the most energetic radio jets, such as those observed in the galaxy cluster MS0735.6+7421. Despite evidence of chaotic accretion in the current study, the jet power may indicate a potential historical phase of spin-up or a scenario where high spin persists even under chaotic accretion conditions. Future studies that integrate spin evolution models with jet energetics could provide deeper insights into this complex interplay.

\subsubsection{Comparison with other tracers}\label{sec:other_tracers}

Over the past three decades, researchers have mapped the kinematics of gas or stars in the nuclear regions of around 150 SMBHs, enabling SMBH mass measurements in systems with ordered and rotating flows \citep[see review by][]{Kormendy2013}. Of these, a few are located at the centers of galaxy clusters \citep[e.g.,][]{McConnell2012,Rusli2013,Thomas2016}; however, most are found in disturbed galaxy clusters, such as NGC 3842, the BCG of Abell 1367, and NGC 4889, the BCG of the Coma cluster. Few SMBHs in relaxed galaxy clusters, where active feedback from the AGN typically occurs, have been studied. Notable exceptions include M87 \citep[][]{Walsh,Osorno2023}, M84 \citep{Walsh2010,Bambic2023}, and NGC 1600 \citep{Thomas2016}. Among these, M87 presents unique challenges due to its AGN being much brighter than those in M84 and NGC 1600.

M84 is a massive nearby elliptical galaxy located in the Virgo Cluster. It is not the BCG of the cluster, but rather a satellite galaxy in orbit around the BCG of the Virgo cluster, M87. In \citet[][]{Walsh2010}, the authors carried out comprehensive gas-dynamical modeling for the emission-line disk within $\sim70$ pc of the nucleus, with the goal of resolving a discrepancy in the literature. In comparison to two previous studies of M84, their analysis included a more thorough treatment of the propagation of emission-line profiles through the telescope and STIS optics, and also accounted for the effects of intrinsic velocity dispersion within the emission-line disk. Similar to our observations of \zn\/, three STIS slits were placed to cover the central AGN in M84. From the STIS kinematics of the [N \textsc{ii}] $\lambda 6583$ emission line in M84, the authors observed clear evidence of rotation, in contrast to what we see in \zn\/. By modelling the M84 emission-line gas as a dynamically cold, thin disk in circular rotation, they determined a best-fit black hole mass of $(4.3^{+0.8}_{-0.7}) \times 10^8 \, M_{\odot}$. Incorporating an asymmetric drift correction into the disk model resulted in a revised best-fit black hole mass of $(8.5{\pm0.9}) \times 10^8 \, M_{\odot}$. Additionally, a strong rise in flux and velocity dispersion was observed along the central slit, with the maximum located at the AGN, similar to what we see in \zn. At this point, the velocity dispersion reached a peak value of $\sigma \sim 400 \, \text{km s}^{-1}$ in M84 (compared to $\sim595$ km s$^{-1}$ in the case of \zn), consistent with the presence of a $\sim 10^8 \, M_{\odot}$ SMBH in M84.

One important difference between \zn\/ and M84 is that, while both exhibit radio-mode AGN feedback, the power of the radio jets in \zn\/, as measured from the mechanical work required to inflate the X-ray cavities \citep[$P_{\rm cav}\sim5\times10^{45}$ erg s$^{-1}$ ][]{Rafferty2006}, is three orders of magnitude greater than that in M84 \citep[][]{Rafferty2006,Russell2013}. The powerful radio jets in \zn\/ may therefore have a stronger influence on the gas flow compared to M84, potentially removing all signs of rotation. 

This is also the case in M87, where the radio jets have a similar power to that of M84, and are thus much less powerful than the ones in \zn\/. As with M84, the ionized gas flow in M87 shows evidence of rotation, allowing for a dynamical mass measurement of the black hole. In \citet[][]{Osorno2023}, the authors used a deeper data set of ionized gas kinematics, obtained with the Multi Unit Spectroscopic Explorer (MUSE) instrument on the Very Large Telescope (VLT), to model the morphology and kinematics of multiple ionized gas emission lines in M87 (primarily H$\alpha$+[N \textsc{ii}] $\lambda\lambda 6548,6583$ and [O \textsc{i}] $\lambda 6300$) in the nucleus of M87. Similarly to M84, the velocity dispersion peaks near the AGN, reaching $\sim400$ km s$^{-1}$. However, the new, deep VLT data revealed complexities in the nuclear ionized gas kinematics that were not apparent in earlier, more limited HST STIS data. Several ionized gas filaments in M87 appear to extend well within the sphere of influence, though they appear to terminate before reaching the black hole. The spectroscopic data also show the presence of an outflow within the sphere of influence in addition to the rotating disk. The complexity of the kinematics makes the accurate measurement of the black hole mass in M87 from the ionized gas kinematics more challenging than initially thought.
Moreover, \citet{Yuzhu2023} analyzed the jet precession using 22 years of radio observation monitoring, linking it to the precession of the accretion flow based on general relativistic magnetohydrodynamics (GRMHD) simulations, which further underscores the complex gas kinematics near the central black hole in M87.

Overall, the fact that we do not see rotation in \zn\/ compared to M84 and M87, may be due to the much more powerful radio jets that might be disrupting the flow of gas.

In the case of \zn\/, the kinematics of the molecular gas phase were also studied using ALMA by \citet[][]{Russell2016}. The authors used the CO(1-0) and CO(3-2) emission lines to trace the kinematics of the cold phase in the gas filaments. The CO(1-0) line traces larger scales, probing several kpcs with a beam size of $1.6 \, \text{arcsec} \times 1.2 \, \text{arcsec}$ (equivalent to $3 \, \text{kpc} \times 2.2 \, \text{kpc}$). In contrast, the CO(3-2) line reaches sub-kpc resolution with a beam size of $0.27 \, \text{arcsec} \times 0.19 \, \text{arcsec}$ (or $500 \, \text{pc} \times 350 \, \text{pc}$). As mentioned in Section \ref{sec:obsAlma}, the CO(3-2) were reanalysed to maximise the spatial resolution and the resulting flux map is shown in Fig. \ref{fig:zoom}. Here, the kinematics of the molecular gas in the filaments show low velocities, remaining within $\pm 100 \, \text{km s}^{-1}$ of the galaxy’s systemic velocity. Additionally, \citet[][]{Russell2016} identified two distinct velocity components when fitting the spectral lines. These components exhibited a FWHM of less than $150 \, \text{km s}^{-1}$, which was significantly lower than the stellar velocity dispersion of the BCG. Overall, \citet[][]{Russell2016} found that the velocity structure of the cold molecular gas is inconsistent with a merger or gravitational free-fall of cooling gas. If the molecular clouds had originated from cooling gas located even a few kpcs away, their velocities would likely be higher than those observed. Instead, the authors suggest that the filaments of cold molecular gas form in the updraft of the rising X-ray cavities.

Interestingly, as shown in Fig. \ref{fig:zoom}, the CO(3-2) map also reveals a lack of emission in the very core of the galaxy. We further discuss this absence of molecular emission in Section \ref{sec:disc_flow_den}.

On another note, by probing the flow of gas within the nuclear regions of \zn, we observe a general positive velocity offset across the area probed by STIS, relative to the reference redshift we used in our study of $z = 0.102428$. This reference redshift was determined by calculating the average redshift on kpc scales using the optical emission line spectra of the ionized gas throughout the BCG \citep{Marie-Joelle}. According to our STIS data, the highest velocity offset occurs at the pixel corresponding to the nucleus (the central pixel in Position 1), where we detect a redshifted velocity of $470 \, \text{km s}^{-1}$. If this central pixel were used as the reference velocity, the corresponding redshift for \zn\/ would be $z = 0.104150$.

Interestingly, similar velocity offsets have been observed in other BCGs, where the kinematics of the gas within the filaments appear to be offset with respect to the kinematics of the BCG itself, sometimes reaching offset velocities of several hundreds of km s$^{-1}$ \citep[][]{Marie-Joelle}. The authors in \citet[][]{Marie-Joelle} suggest that the multiphase gas on kpc scales is being stirred/churned to higher velocity dispersions by the rising X-ray cavities. In the case of \zn\/, the STIS observations indicate that such velocity offsets may also extend down to sub-kpc scales. We further discuss the comparison between the larger-scale kinematics and our STIS results, as well as the specific case of \zn\/ and this churned gas, in greater detail in Section \ref{sec:disc_obs}.

Finally, we note the gradual rise in velocity along Position 1, which increases radially from $171^{+34}_{-36} \, \text{km s}^{-1}$ to $470^{+31}_{-30} \, \text{km s}^{-1}$ at the location of the nucleus. This may suggest that the gas is in free-fall, though it is challenging to confirm. To explore this possibility, we follow the same procedure as \citet[][]{Russell2016} to estimate the final velocity of a free-falling gas blob. We use the same parameters as in their Section 4.2, but we start from a radius of 570 pc and extend down to 0 pc, with an initial velocity of approximately $171 \, \text{km s}^{-1}$. Using this method, we find that the final velocity for a free-falling blob would be around $350$ km s$^{-1}$. It is important to note that this is a rough approximation, as it does not account for the gravitational potential of the SMBH, which would likely result in an even higher velocity. Therefore, while our results are roughly consistent with the possibility of free fall, more detailed modeling is needed to draw a definitive conclusion.

\subsection{Accretion flow: density}\label{sec:disc_flow_den}

M84 is one of the rare systems in which the Bondi radius can be resolved by the $Chandra$ X-ray Observatory. The other four systems are Sgr A$_{\text{*}}$ \citep[e.g.,][]{Baganoff2003}, the nearby galaxy NGC 3115 \citep{Wong2014}, the massive BGG NGC 1600 \citep{Runge2021}, and M87 \citep{Russell2015}. M84 and M87 stand out compared to the other systems, due to their active feedback, which carves out X-ray cavities \citep{Bambic2023}, although three orders of magnitude less powerful than those in \zn. 

In \cite{Bambic2023}, the authors analyzed deep $Chandra$ observations of M84 (840 ks) and found that the accretion flow does not follow the expected behaviour from Bondi accretion and is instead strongly influenced by the AGN’s bipolar radio jets. Along the jet axis, the density profile is consistent with r$^{-1}$ on 100 pc scales, but the profile becomes flatter perpendicular to the jet. The density profiles they found were overall completely inconsistent with the profile expected by the \citet{Bondi1952} solution for an adiabatic gas at more than a 5$\sigma$ level. 

This is similarly the case in the massive BGG NGC 1600, although less powerful feedback is occurring in this system compared to M84. Here, the density slope is also flat within the Bondi radius, consistent with r$^{-0.61}$ on scales of 300 to 500 pc. \citet[][]{Russell2018} also found a similar result for M87.

\begin{figure}
\centering
\includegraphics[width = 1\columnwidth]{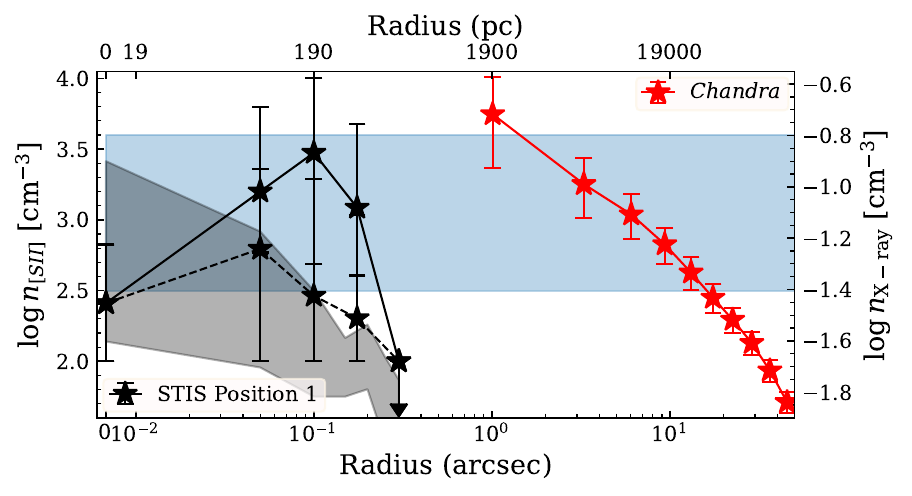}
\caption{Density estimated from the ratio of the \sii{} emission lines (black) for the ionized gas along the slit Position 1, as well as from the ICM electrons from projected \textit{Chandra} X-ray observations (red).
The northeast offset (radius) for STIS observations is plotted by the solid line, and the dashed line represents the offset towards the southwest direction from the central AGN.
The density near the vicinity of the central SMBH ($r\lesssim600$ pc) shows a notably flat distribution within the errors. 
The shaded region in black shows the estimate of the density of cold gas measured from MHD simulations.
The shaded region in blue simply highlights the extent of the flat profile throughout the core, in both ionized and X-ray gas.}
\label{fig:density_stis_xray}
\end{figure}

In \zn\/, the density profile of the ionized gas traced by the \sii{} emission lines observed with STIS (see Fig. \ref{fig:density_stis_xray}) is also remarkably flat across all three slit positions. This indicates that the ionized gas exhibits a similarly flat density profile on comparable scales, around 100 pc, as seen in the hot X-ray gas of M84 and NGC 1600. Indeed, if the ionized gas originates from cooling of the hot ICM, the two phases could be connected. 

The flat density profile can be attributed to the presence of a very large SMBH at the center (see Section \ref{sec:disc_BHmass}). As galaxies merge in dense environments, their central SMBHs can form a binary system. Interactions between this binary and the surrounding stars can eject stars from the core, leading to a depleted stellar density—a phenomenon known as core scouring. This process facilitates the merger of the SMBHs, resulting in a more massive central SMBH \citep[see also][]{Natarajan2009,Armitage+2002}. Consequently, flat stellar surface brightness profiles are also predicted, such as the one observed here in \zn \citep{Nasim+2021}. The flat density profile (see shaded region in blue in Fig. \ref{fig:density_stis_xray}) might also directly result from such scouring. However, more detailed calculations are necessary to better understand the role of binary SMBHs and their impact on gas versus stars.

In Fig. \ref{fig:zoom}, we also observe a gap in the CO(3-2) emission near the central regions, coinciding with where we are probing the dynamics of the ionized gas using the STIS observations. While some BCGs have shown cold molecular gas absorption along the line of sight to the AGN \citep[][]{Rose2019b}, an examination of the ALMA spectrum reveals no evidence of absorption in the core of our system. Interestingly, a similar CO(3-2) void has been noted in other BCGs. This gap could indicate the need for a denser gas tracer in this region, despite the flat density profile of the ionized gas observed in our STIS data. Alternatively, it could be the result of a torus structure. The lack of molecular gas could also be driven by the AGN, although we don't see any evidence of outflow in the molecular gas, see also \citet{Garcia2014}. Further observations are necessary to better understand this phenomenon. In Section \ref{sec:disc_sim}, we compare the density slope measured with STIS to predictions from state-of-the-art simulations.

\citet[][]{Bambic2023} also found that the accretion rate estimated from Bondi flow in M84 is inconsistent with the rate required to explain the observed feedback. In this work, the authors estimate the accretion rate from Bondi flow down to the innermost stable orbit (ISCO) of the SMBH, using the radial dependence of accretion in the radiatively inefficient accretion mode as described by \citep[][]{Ressler2018,Ressler2020}, which traces the flow from large scales down to the black hole horizon. Their model predicts that the suppressed Bondi accretion rate down to the ISCO follows the scaling $\dot{M}=(R_{\rm ISCO}/R_{\rm B})^{1/2}\dot{M}_{\rm B}$. Using this, \citet[][]{Bambic2023} estimate the accretion rate at the ISCO to be $\sim 8.5\times10^{-6}$ M$_\odot$ yr$^{-1}$ in M84. Assuming a 10 per cent efficiency, the inferred power that the SMBH can release is $L_{\rm ISCO}\sim5\times10^{40}$ erg/s, which falls short of the jet power of $1.1^{+0.9}_{-0.4}\times10^{42}$ erg/s. A similar discrepancy is noted for NGC 1600 \citep{Runge2021}. Consequently, additional sources of gas, such as cold gas, may be required to sustain SMBH feeding. While this interpretation is correct, it is important to emphasize that the inferred power depends on the specific scaling of the accretion rate with radius and assumes a 10 per cent jet efficiency. GRMHD simulations suggest that jet efficiencies can be much higher, even exceeding unity, depending on the black hole spin and magnetic field configuration \citep[e.g.][]{Techkhovskoy2010}.


Here, we are unable to resolve the Bondi radius using X-ray observations. The Bondi radius ($r_B$) is defined as:

\begin{equation}
\frac{r_B}{\text{kpc}} = 0.031 \left( \frac{k_B T}{\text{keV}} \right)^{-1} \left( \frac{M_{BH}}{10^9 M_\odot} \right),
\end{equation}\label{eq:rb}

where $k_B T$ is the temperature in the core. To obtain a rough estimate of the Bondi radius, we used the innermost bin of the temperature profile from \citet{Sanders2014}, where $k_B T \sim 2.5$ keV within the inner 10 kpc. Additionally, we adopted the SMBH mass determined in Section \ref{sec:disc_BHmass} as $1.5 \times 10^{10} M_\odot$. This yields a Bondi radius of approximately $r_B \sim 180$ pc, or roughly $0.1\arcsec$. The pixel size of 0.49\arcsec for $Chandra$ is therefore unable to resolve the Bondi radius in this source at X-ray wavelengths, which is not surprising given its redshift. However, if there were to be significant cooling, such as a large cooling flow (see Section \ref{sec:disc_fut}), the Bondi radius could be much larger. In Fig. \ref{fig:density_stis_xray}, we also compare the density profile of the ionized gas to the electron density profile of the X-ray gas. 

In \citet[][]{Bambic2023}, the authors argued that on large scales in BCGs and galaxies like M84, feedback from radio jets prevent hot gas from cooling over distances of tens of kpc. However, on smaller scales—particularly near the Bondi radius, where the cooling time drops below 0.1 Gyr—these heating mechanisms become less efficient. They suggest that the accretion flow near the Bondi radius should not be viewed as a static equilibrium between AGN heating and radiative cooling but rather as a dynamic flow influenced by the SMBH. We may be observing a similar process here in \zn: while feedback effectively thermalizes most of the hot gas on large scales, within the core, the flow decouples, resulting in a more chaotic flow with lower angular momentum, which may lead to a self-regulating feedback loop.

Supporting this, \citet[][]{Russell2018} found that the lowest temperature and most rapidly cooling gas in M87 is located on scales of around 100 pc, where a mini cooling flow may be forming. This flow could feed the cold gas disc around the nucleus of M87. The disc itself has a radius of approximately 1\arcsec (80 pc), positioning it just within the transition region in the X-ray gas properties observed at the Bondi radius ($80-250$ pc). Based on their estimates, the mini cooling flow could supply the cold gas in the disc on timescales of a few $10^7-10^8$ years, although some of this cooling gas may be transformed into star formation before it reaches the nucleus \citep[][]{McNamara1989}.

\subsection{Accretion flow: ionization} \label{sec:disc_flow_ion}

The flux ratios of optical emission lines from extragalactic sources provide valuable diagnostics for probing the excitation conditions of ionized nebular gas \citep[e.g.,][]{Baldwin1981,Kewley2006,Smirnova2007}.
For example, the \nii{} to H$_\alpha$ ratio is sensitive to the hardness of the illuminating SED \citep[AGN or star formation; e.g.,][]{Bae2017} and other various excitation mechanisms that can ionize the gas, such as shocks \citep[e.g.,][]{Singh2013,Heckman1980,Allen2008}.
The ratio of \nii{}/H$_\alpha\gtrsim0.6-1$ is typically an indication that the gas is ionized by a continuum source with a harder UV flux than a starburst.
In clusters, collisional heating and emission from hot ICM have been suggested as additional mechanisms that can potentially enhance the \nii{}/H$_\alpha$ ratio \citep{Ferland2009,Fabian2011}.

\citet{Hamer2016} presented a survey of 73 galaxy clusters and groups with the \textit{VLT-VMOS}, where they investigated the excitation states of the nebular gas in BCGs.
They found an overall trend with the \nii{}/H$_\alpha$ ratio decreasing as the distance from the nuclear increases (a few tens of kpc),
indicating a harder radiation field near the nucleus.
\citet{Tamhane2022} observed the BCG in the Abell 1795 cluster with the \textit{VLT-MUSE} and reported a similar trend in which the \nii{}/H$_\alpha$ ratio sharply drops from $\sim0.85$ to $\lesssim0.7$ between the nuclear regions and $\sim4$ kpc away from the nucleus.
While it is challenging to pinpoint what is the main driver that determines the range of line ratios observed in the BCGs, since both AGN and star formation activities are occurring in many of these galaxies, what has been consistently found is the strong evidence that AGN activity predominantly drives ionization near the center.

In \zn\/, we found that the \nii{}/H$_\alpha$ ratio remains greater than $\sim1$ throughout the inner $\sim600$ pc region with no clear trend of decreasing with increasing distance from the nucleus (see Figs.~\ref{fig:nhrat_flux} and \ref{fig:3d_ratio_maps}).
This means that if we assume that \zn\/ also has a similar large-scale trend of increasing ratio towards the center, our results show that the ratio saturates to $\sim1-1.5$ and remains constant in the nuclear region within the errors.
The consistently high \nii{}/H$_\alpha$ ratios observed throughout the STIS-covered region suggest that AGN radiation, and potentially shock ionization, are the primary contributors to ionization extending out to a few hundred pcs, similar to those found in earlier work.
Considering that the ionizing flux from the AGN decreases $\propto r^2$, we speculate that the fraction of contribution from the shock becomes more significant with distance for ionizing the gas in the vicinity of the nucleus out to hundreds of parsecs to maintain a similar level of excitation conditions. 
Although the evidence is subtle, we observe a weak trend of elevated \nii{}/H$_\alpha$ ratios toward the X-ray cavities along the north-south axis (see Fig.~\ref{fig:3d_ratio_maps}).
This observed pattern may be capturing the link between AGN activity and the rising cavities, which are often traced by radio emission \citep[e.g.,][]{McNamara2009}.




\subsection{Connecting gas flows to the larger-scale: observations} \label{sec:disc_obs}

As mentioned in the introduction, the large-scale kinematics and morphology of ionized and neutral ISM in \zn\/ have been studied previously with ground-based integral field spectrographs (IFSs). The spatially resolved kinematic properties of the Pa$\alpha$ and ro-vibrational H$_2$ lines have been investigated using \textit{VLT-SINFONI} near-infrared IFS \citep{Wilman2009}. They were able to identify extended line emission features ($\sim 5$ kpc long) and presented the gradual velocity shifts in the kinematic maps as evidence for the coordinated rotational motion of the ISM around the nucleus. The velocity dispersion maps for both Pa$\alpha$ and H$_2$ lines showed peaks in the nuclear region, consistent with the patterns observed in the optical nebular emission lines (\S~\ref{sec:reskin}).

\citet{Marie-Joelle} observed \zn\/ with Keck Cosmic Web Imager (KCWI) to study the large-scale kinematic properties of nebular gas in the vicinity of AGN using [\ion{O}{2}] emission maps. Significantly larger nebular emission features ($\sim10-20$ kpc) than those reported in \citet{Wilman2009} have been detected from the [\ion{O}{2}] emission line, allowing them to investigate the large-scale kinetic properties of the warm ionized gas around the BCG.
They reported evidence of churned (or disrupted) gas near the nucleus, traced by high-velocity dispersions extending out to a few kpcs, and suggested that the radio AGN activity is the likely cause of the observed kinematic disturbances and outflows. 
Notably, their emission maps showed overall positive velocity offsets relative to the systematic redshift of the BCG, with no robust signatures of rotation-dominated gas motion.
They postulated that the proper motion of the central galaxy relative to the hot atmosphere toward the observer along the line of sight produces such a systematic shift in the kinematic structure. One possible reason for the difference is that \citet{Wilman2009} probed gas kinematics on much smaller scales with higher spatial resolution, potentially capturing finer details in the velocity structure. Moreover, the differences in the kinematic properties traced by the emission lines from recombination and rotation–vibration transitions of hydrogen and [\ion{O}{2}] may arise from the fact that they probe different phases of the ISM with different physical properties (e.g., density). The nebular lines are better probes of the kinematics of diffuse gas, which are more susceptible to the influence of the central AGN \citep[e.g.,][]{Hatch2007}.

Despite the dramatic differences in the size scales of the ISM probed by KCWI (and \textit{SINFONI}) and STIS, we also observed similar kinematic properties to those found in a larger-scale nebular gas. This result suggests that the outflows and disruption in the ISM caused by AGN activity originate from much smaller scales. The velocities and the velocity dispersions we measured in the central pixels of the STIS observation are $\sim2-3$ times greater than the values reported by \citet{Marie-Joelle}. We also find positive velocities (with respect to the galaxy; \S~\ref{sec:reskin}), suggesting that the nebular gas is moving away from the galaxy along the line of sight, as seen in large-scale kinematics in KCWI data.
However, we do not see the complex velocity function due to the proper motion of the galaxy discovered in \citet{Marie-Joelle} (i.e., a V-shaped velocity gradient) from the STIS data, likely because the kinematics of the nebular gas close to the SMBH is more dominated by the AGN activity and chaotic flows than the large scale motion. Considering the spatial resolution of KCWI ($\sim1\arcsec$), the complex kinematics of the nebular gas within the central $\sim600$ pc probed by STIS are averaged in a single central pixel in the KCWI emission maps, which may result in such differences in the kinematic properties.

Nevertheless, the similarities between the small-scale ISM features and larger-scale properties, such as the increase in velocity dispersion towards the center, indicate that kinetic flow within the nuclear regions may still be coupled with the larger reservoir of nebular gas. 

The kinematics of large-scale cold molecular gas ($\sim3-5$ kpc) in \zn\/ have been traced using CO emission lines, revealing significantly lower FWHM ($\lesssim150\ \mathrm{km\ s^{-1}}$) compared to those observed in warm ionized gas \citep[\S~\ref{sec:other_tracers};][]{Russell2016}.
An investigation was conducted to compare the kinematic properties of cold molecular gas and warm ionized nebular gas traced by [OII], and no conclusive evidence supporting the physical relation between the two phases of the gas was discovered \citep{Marie-Joelle}.
No obvious correlation is observed between the kinematics of large-scale cold molecular gas and the ionized gas within the sphere of influence of the central SMBH probed by STIS.


\subsection{Connecting the gas flow to the larger-scale: simulations}\label{sec:disc_sim}

\begin{figure}
\centering
\includegraphics[width = 1\columnwidth]{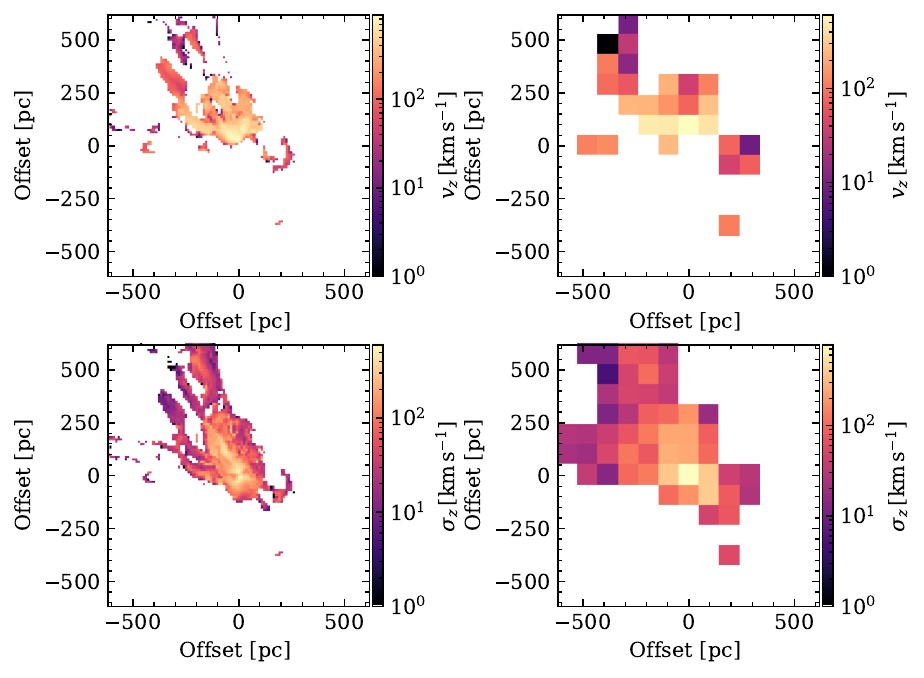}
\caption{Velocity along the line of sight (top), and velocity dispersion (bottom) measured from MHD simulations with full resolution (left) and coarse-grained resolution similar to the observations (right). The turbulent structure of the accretion flow is illustrated.}
\label{fig:sims}
\end{figure}

Our results show that the cold gas dynamics share many similarities with the chaotic accretion flow characterized in \citet{Guo2024ApJ...973..141G}. Therefore, in this section, we compare our STIS-measured flow results with predictions from 3D magnetohydrodynamic (MHD) simulations of SMBH fueling from a turbulent cooling medium on galactic scales by \citet[][]{Guo2024ApJ...973..141G}. Bridging the very same scales explored here, \citet[][]{Cho2023, Cho2024} also track both feeding and feedback self-consistently in GRMHD simulations of adiabatic Bondi-like accretion. These simulation suites are among the first to trace accretion flow from kpc scales down to the horizon scale of SMBHs. 

The simulations deployed here are similar to those of \citet{Guo2024ApJ...973..141G}, but use a larger black hole mass of $1.5 \times 10^{10}M_\odot$, based on the rough estimate in this work. Details of the methodology can be found in \citet{Guo2023ApJ...946...26G} and \citet{Guo2024ApJ...973..141G}. We then measured the properties of the accretion flow that emerged in these new MHD simulations.

In Fig. \ref{fig:sims}, we show resulting images of line-of-sight velocity and velocity dispersion of the gas at $T\sim 10^4,\mathrm{K}$, displayed at full spatial resolution as well as at a coarse-grained resolution comparable to the STIS observations ($95$ pc). These images reveal the chaotic structure of the gas flows, where neighboring pixels display velocity differences of $\pm100-300$ km s$^{-1}$. The velocity dispersion also increases toward the SMBH, similar to what is observed with our STIS observations. We note however that Fig. \ref{fig:sims} represents a single snapshot of the gas flows (i.e., a single realization). The fact that the observations do not show evidence of rotation is also consistent with the simulation as the size of the rotationally supported gas flow typically only spans several tens of parsecs in simulations, smaller than the resolution of the current observations. 


In Fig. \ref{fig:density_stis_xray}, we also plot the distribution of the gas density measured at a resolution of 95 pc in a series of snapshots covering 50 Myr from the simulation and compare it with the observations. The gas density in the simulations, while slightly lower by a factor of a few, is overall similar to observations. 

Hence, these results suggest that cooling flows from large-scale hot gas, combined with magnetic fields and large-scale turbulence, can spontaneously generate small-scale chaotic cold flows and density structures similar to those seen in observations. 

As presented by many simulations~\citep[e.g.,][]{Ressler2018, Ressler2020, Guo2023ApJ...946...26G, Xu2023}, the hot gas density scales with radius as $\rho\propto r^{-1}$, which gives a density of $\sim 1-10 \,\mathrm{cm^{-3}}$ on scales of $\sim 100 \,\mathrm{pc}$ insufficient to explain the ionized gas density of $\sim 10^3\,\mathrm{cm^{-3}}$ estimated from the observations. This again indicates that cold gas is required to predict ionized gas density. 
Efforts to model cold gas around SMBH from realistic environments also exist in \citet{Kaaz2024}, where they performed GRMHD simulations of highly magnetized quasar disks around an SMBH of mass $M\approx10^7 M_\odot$ from cosmological initial conditions.
However, the density of the cold ionized gas is harder to predict due to its thermodynamics and its complex dependence on magnetic field structure. For example, the magnetic field may vertically support the disk, leading to a smaller density.

We also note that the gas flow in our simulations is highly chaotic and varies over time, making precise comparisons challenging. Furthermore, the thermodynamics of the cold gas are not fully reliable, as our resolution is insufficient to capture the scale of cold, dense gas. We therefore leave a detailed comparison of the density for future work. Indeed, upcoming simulations with improved resolution and more realistic feedback may help us better understand the gas flow, and in particular, the density structure. Rapid progress is currently underway exploring this complex problem using GRMHD simulations that couple scales - from pc to kpc and beyond - and we anticipate being able to make even more meaningful comparisons with data in the very near future.

\subsection{Future observations}\label{sec:disc_fut}

The HST STIS observations presented certain limitations. We detected emission only within the central pixels, and beyond a radius of 570 pc, the signal became too faint to be captured, even by HST. Additionally, the long slit configuration probed just a single direction, restricting the spatial information. In contrast, the James Webb Space Telescope (JWST) offers the potential for full integral field unit (IFU) observations in the near-infrared and infrared bands, which would capture a broader range of emission lines, enabling a deeper analysis of the nuclear gas conditions.

JWST's NIRSpec wavelength coverage further enhances these capabilities, providing a rich set of diagnostic lines, including molecular H$_2$ transitions, which we know are in \zn\/ \citep[][]{Edge2002,Wilman2009}, as well as coronal lines from highly ionized species like [{\sc Si\,VI}] and [{\sc Mg\,VIII}]. The H$_2$ 1-0 S(3) transition in particular matches the morphology of the ALMA CO(3-2) gas. The ratios of these lines can also help us constrain shock contributions linked to AGN feedback and indirectly probe the intrinsic shape of the AGN's spectral energy distribution, offering further insights into the accretion properties of the central SMBH.

Additionally, recent work by \citet[][]{FabianHCF1,FabianHCF2,FabianHCF3,FabianHFC4} suggests that significant cooling of the hot ICM still occurs in galaxy clusters, even in the presence of powerful AGN feedback. They propose a "Hidden Cooling Flow" model, where much of the ICM cools but is obscured by cold clouds and dust near the cluster centers. Cooling rates in this model range from $15-50$ per cent of the predicted values, with some clusters exhibiting rates as high as several hundred solar masses per year. Although AGN feedback moderates these rates, these authors suggest that it does not fully prevent cooling and that cooled gas may still contribute to star formation or accretion by central SMBH. If these cooling flows exist, they should re-emerge in the far-infrared, and future JWST observations could reveal this critical phase of gas cooling and SMBH accretion.

\begin{figure}
\centering
\includegraphics[width = 1\columnwidth]{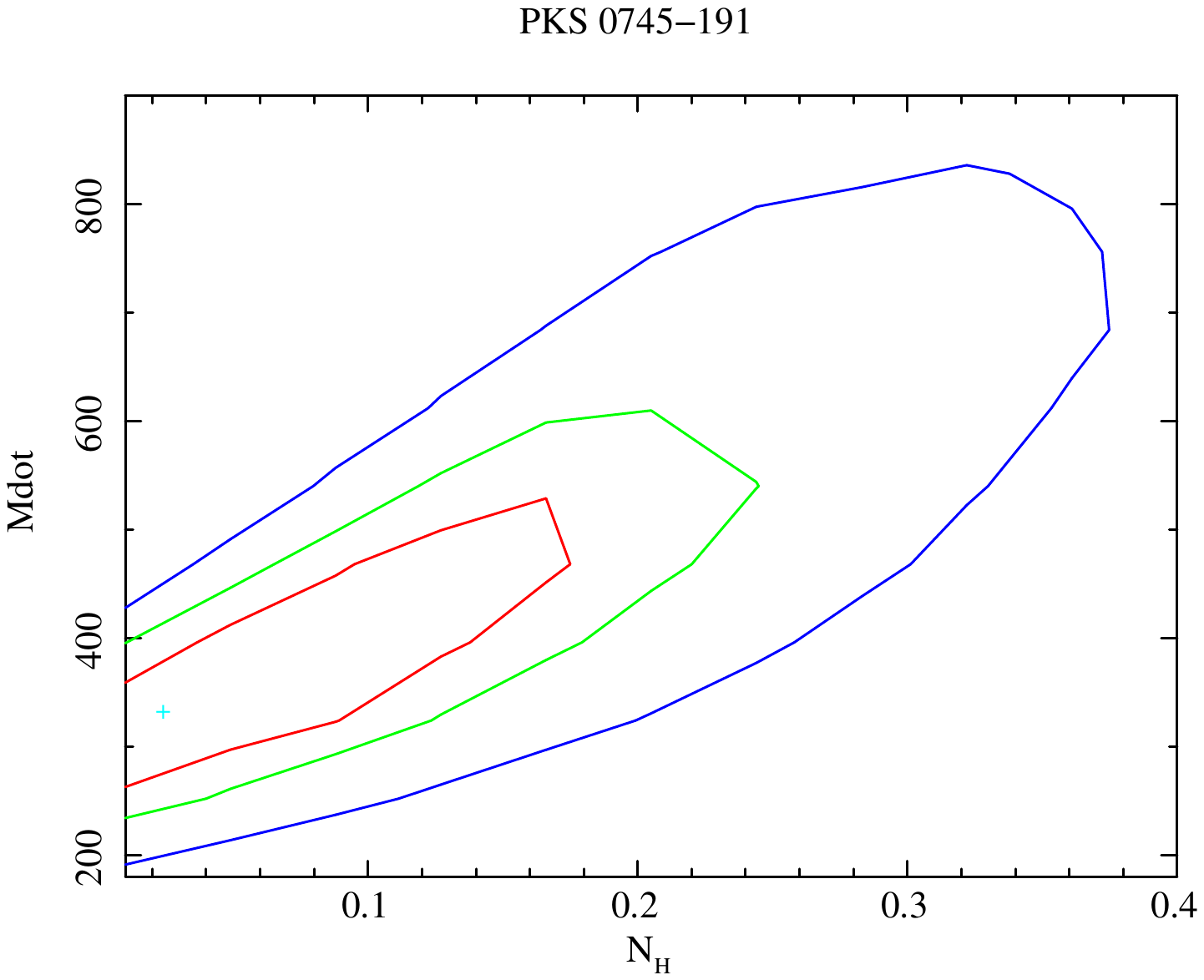}
\caption{The mass cooling rate (Mdot) and intrinsic column density ($N_{\rm H}$), expressed in units of $10^{22}$ cm$^{-2}$, were derived from a Hidden Cooling Flow model fit to the XMM RGS spectrum of \zn. This analysis reveals that the XMM RGS data support the presence of substantial cooling within the system. Specifically, the inferred mass cooling rate is Mdot $= 332^{+88}_{-80}$ M$_\odot$ yr$^{-1}$, indicating active and significant cooling of gas at the core of \zn. A Galactic column density of $4.2\times 10^{21}$ cm$^{-2}$  has been applied \citep{Sanders2014}.}
\label{fig:RGS}
\end{figure}

We have analysed the archival XMM-Newton RGS observations of PKS 0745-191 with a total expsoure of  281 ks  summing both spectrometers. An intrinsic absorption model was used in which the absorbing matter and emitting gas are interleaved \citep{FabianHFC4}. The results are shown in Fig. \ref{fig:RGS}. The mass cooling rate $\dot M = 332^{+88}_{-80}$ M$_\odot$ yr$^{-1}$, which is consistent with significant cooling of gas in the core. These findings provide further evidence that AGN feedback, while impactful, may not completely suppress cooling flows in galaxy clusters such as \zn\/, allowing a portion of the cooled gas to potentially fuel star formation or accretion onto the central SMBH.

\section{Conclusions} \label{sec:conc}

In this paper, we present detailed kinematic and density observations within the core of \zn, which is undergoing powerful radio-mode AGN feedback. By using high-resolution STIS data from the HST, we mapped the ionized gas around and within the sphere of influence of the SMBH in \zn, investigating how powerful relativistic jets impact the surrounding intracluster medium (ICM) on sub-kpc scales. Our observations provide key insights into the nature of AGN feedback and its role in the evolution of massive galaxies.

First, based on the velocity dispersion of the ionized gas increasing sharply at the location of the SMBH, we estimated the SMBH mass to be $\sim1.5\times10^{10}\,\rm M_\odot$, making the SMBH in \zn\/ potentially one of the most massive known.

Second, our observations suggest that the flow of ionized gas on sub-kpc scales, well within the SMBH's sphere of influence, is highly chaotic. This lack of organized rotation, likely due to the disruptive effects of the powerful radio jets, makes it difficult to obtain a precise dynamical mass measurement of the SMBH. In environments with intense radio-mode AGN feedback, these jets may reduce angular momentum within the gas, which can enable accretion to occur on very small scales. This behavior points to a self-regulating feedback mechanism: while the jets inhibit large-scale cooling, they do not entirely prevent accretion in the core, where chaotic gas motions can still dominate. Compared to other BCGs with weaker jets, where rotational structures are more stable, the gas in \zn\ appears significantly more turbulent, suggesting a stronger coupling between jet activity and gas dynamics on these smaller scales.

Third, our density analysis revealed a notably flat profile in the ionized gas, comparable to the flat profiles observed in the X-ray gas around other galaxies where the Bondi radius is resolved.

Fourth, by comparing the observed kinematics with studies of large-scale ionized gas, we observed that AGN-driven turbulence extends across several kpcs. This suggests a coupling between small-scale nuclear dynamics and larger-scale ICM processes.
We note an overall increase in gas velocity and velocity dispersion toward the nucleus, along with large velocity jumps within the sphere of influence, consistent with the presence of churned gas extending down to the core.

Finally, the observed \nii{}/H$_\alpha$ ratios seem to indicate AGN ionization extending hundreds of parsecs from the nucleus, with a potential contribution from shock ionization, particularly along the radio jet axis.

Overall, our findings indicate that sub-kpc scale flows in systems with powerful radio-mode feedback are highly chaotic, and that this may lead to a self-regulating feedback loop. On larger scales, despite strong feedback, \zn\ continues to fuel star formation and maintain a substantial molecular gas reservoir, demonstrating that AGN feedback does not entirely prevent cooling in massive clusters. Future JWST observations will enable a deeper exploration of these accretion processes through emission lines such as H$_2$ 1-0 S(3), and may reveal hidden cooling flows that contribute to SMBH growth in cluster environments, both of which can be probed with instruments on JWST like NIRSpec.

\begin{acknowledgments}

We thank Nicholas McConnell for his invaluable guidance in setting up the HST observations, which greatly contributed to the success of this work. JHL acknowledges funding support from the Canada Research Chairs Program, as well as the Natural Sciences and Engineering Research Council of Canada (NSERC) through the Discovery Grant and Accelerator Supplement programs. MP acknowledges funding from the Physics department of the University of Montreal (UdeM) and the Centre for Research in Astrophysics of Quebec (CRAQ). ARL is supported by the Gates Cambridge Scholarship, by the St John's College Benefactors Scholarships, by NSERC through the Postgraduate Scholarship-Doctoral Program (PGS D) under grant PGSD3-535124-2019 and by FRQNT through the FRQNT Graduate Studies Research Scholarship - Doctoral level under grant \#274532. ACE acknowledges support from STFC grant ST/T000244/1. PN acknowledges support from the Gordon and Betty Moore Foundation and the John Templeton Foundation that fund the Black Hole Initiative (BHI) at Harvard University where she serves as one of the PIs. MLGM acknowledges financial support from NSERC via the Discovery grant program and the Canada Research Chair program.

Some/all of the data presented in this paper were obtained from the Mikulski Archive for Space Telescopes (MAST). STScI is operated by the Association of Universities for Research in Astronomy, Inc., under NASA contract NAS5-26555. Support for MAST for non-HST data is provided by the NASA Office of Space Science via grant NNX13AC07G and by other grants and contracts. Support for program 14669 was provided by NASA through a grant from the Space Telescope Science Institute, which is operated by the Associations of Universities for Research in Astronomy, Incorporated, under NASA contract NAS5-26555.
The specific observations analyzed can be accessed via doi:\href{http://doi.org/10.17909/tgnq-sp49}{10.17909/tgnq-sp49}.

The authors are pleased to acknowledge that the simulation reported on in this paper was substantially performed using the Princeton Research Computing resources at Princeton University, which is consortium of groups led by the Princeton Institute for Computational Science and Engineering (PICSciE) and Office of Information Technology's Research Computing.

The Universit\'e de Montr\'eal recognizes that it is located on unceded (no treaty) Indigenous territory, and wishes to salute those who, since time immemorial, have been its traditional custodians. The University expresses its respect for the contribution of Indigenous peoples to the culture of societies here and around the world. The Universit\'e de Montr\'eal is located where, long before French settlement, various Indigenous peoples interacted with one another. We wish to pay tribute to these Indigenous peoples, to their descendants, and to the spirit of fraternity that presided over the signing in 1701 of the Great Peace of Montr\'eal, a peace treaty founding lasting peaceful relations between France, its Indigenous allies and the Haudenosauni Confederacy (pronounced: O-di-no-sho-ni). The spirit of fraternity that inspired this treaty is a model for our academic community.

\end{acknowledgments}

%

\facilities{HST (ACS, WFC3, STIS), MAST (HAP), Chandra, ALMA}

\bibliography{main}{}
\bibliographystyle{aasjournal}



\end{CJK}
\end{document}